%% file: main.tex
\documentclass[twoside,leqno,twocolumn]{article}

\usepackage[letterpaper]{geometry}

\usepackage{ltexpprt}
\usepackage{amsmath}
\usepackage{tabularx,booktabs}
\usepackage{authblk}
\usepackage{xspace}
\usepackage{hyperref}
\usepackage{xcolor}

\usepackage{float}
\usepackage{balance}

\newtheorem{definition}{Definition}

\usepackage{graphicx}
\graphicspath{{./figures/}}

\input{./notation-defs.tex}

\begin{document}

\title{\Large Approximating \textsc{Vertex Cover} using Structural Rounding\thanks{This research was supported in part by the Army Research Office under Grant Number W911NF-17-1-0271 to Blair D. Sullivan and the Gordon \& Betty Moore Foundation’s Data-Driven Discovery Initiative under Grant GBMF4560 to Blair D. Sullivan. The views and conclusions contained in this document are those of the authors and should not be interpreted as representing the official policies, either expressed or implied, of the Army Research Office or the U.S. Government. The U.S. Government is authorized to reproduce and distribute reprints for Government purposes notwithstanding any copyright notation herein.}}

\author[1]{Brian Lavallee}
\author[2]{Hayley Russell}
\author[1]{Blair D. Sullivan}
\author[2]{Andrew van der Poel}

\affil[1]{University of Utah, Salt Lake City, UT, USA}
\affil[2]{North Carolina State University, Raleigh, NC, USA}

\date{}

\maketitle







\begin{abstract} \small\baselineskip=9pt

In this work, we provide the first practical evaluation of the structural rounding framework for approximation algorithms. Structural rounding works by first editing to a well-structured class, efficiently solving the edited instance, and ``lifting" the partial solution to recover an approximation on the input. We focus on the well-studied \textsc{Vertex Cover} problem, and edit to the class of bipartite graphs (where \textsc{Vertex Cover} has an exact polynomial time algorithm). In addition to the \naive lifting strategy for \textsc{Vertex Cover} described by Demaine et al. in the paper describing structural rounding, we introduce a suite of new lifting strategies and measure their effectiveness on a large corpus of synthetic graphs. We find that in this setting, structural rounding significantly outperforms standard 2-approximations. Further, simpler lifting strategies are extremely competitive with the more sophisticated approaches. The implementations are available as an open-source Python package, and all experiments are replicable.

\end{abstract}

\section{Introduction}
\input{./sections/intro.tex}


\section{Preliminaries}\label{sec:prelims}
\input{./sections/prelims.tex}

\section{Lifting Algorithms}\label{sec:algs}
\input{./sections/algs.tex}

\section{Experimental Setup}\label{sec:setup}
\input{./sections/setup.tex}

\section{Implementation} \label{sec:implementation}
\input{./sections/implementation.tex}

\section{Experimental Evaluation} \label{sec:experiments}
\input{./sections/experiments.tex}

\section{Conclusion} \label{sec:conclusion}
\input{./sections/conclusion.tex}

%
%
%
%

%
\bibliographystyle{splncs04}

\input{./sections/bibliography.tex}
 \clearpage
 \onecolumn
 \appendix
%

\input{./sections/app-1.tex}

\end{document}

%% file: notation-defs.tex
\newcommand{\kPsieditsG}{\psi_k(\psi_{k-1}( \cdots \psi_2(\psi_1(G))\cdots))}

\def\mC{\mathcal{C}}
\def\mCp{\mathcal{C}_\lambda}

\newcommand{\optsol}[2]{\opt_{#1}(#2)}
\DeclareMathOperator*{\optop}{OPT}
\newcommand{\opt}{\ensuremath{\optop}}
\DeclareMathOperator*{\costop}{Cost}
\newcommand{\cost}{{\ensuremath{\costop}}}

\newcommand{\Problem}[1]{\textsc{#1}\xspace}

\newcommand{\Editfullp}{\Problem{$(\mathcal{C}_\lambda, \psi)$-Edit}}
\newcommand{\partialsol}{S'}
\newcommand{\editgraph}{G'}
\newcommand{\editset}{X}
\newcommand{\liftsol}{L}
\newcommand{\leftover}{I}
\newcommand{\liftsolone}{L_1}
\newcommand{\liftsoltwo}{L_2}
\newcommand{\optlift}{L^*}
\newcommand{\naive}{na\"ive\xspace}
\newcommand{\Naive}{Na\"ive\xspace}
\newcommand{\editfirst}{\texttt{oct-first}\xspace}
\newcommand{\bipfirst}{\texttt{bip-first}\xspace}
\newcommand{\rec}{\texttt{recursive}\xspace}
\newcommand{\recedit}{\texttt{recursive-oct}\xspace}
\newcommand{\recbip}{\texttt{recursive-bip}\xspace}

\newcommand{\vc}{\textsc{Vertex Cover}\xspace}

%% file: sections/intro.tex
Approximation algorithms are a crucial tool for navigating the tradeoff between solution quality and runtime, yet
many approximation algorithms' quality guarantees rely on the structure of limited classes of graphs, undercutting the scope of their applicability. \emph{Structural rounding} is a recent framework~\cite{DEMAINE} for extending approximation algorithms for restricted classes to graphs ``near'' those classes. The approach consists of three phases: editing the input graph to have a given structural graph property (e.g. bounded treewidth), efficiently solving the edited-to instance, then ``lifting" the partial solution to a valid outcome on the original graph.
In this work, we provide the first implementation and experimental evaluation of structural rounding, using it to solve \vc on near-bipartite graphs.

We chose this problem-graph class pair because of the famous result of K{\"o}nig, that \vc is polynomial time solvable on bipartite graphs~\cite{KONIG}. This can be achieved via the Hopcroft-Karp algorithm~\cite{HOPCROFT}. The theoretical guarantee of structural rounding in this setting is considerably worse than that of the standard $2$-approximations (see Section~\ref{sec:prelims-sr}). However, we show experimentally that structural rounding is in fact competitive with (and often much better than) the traditional approaches despite having weaker theoretical guarantees.

While many algorithms already exist for the first two phases of structural rounding, the notion of ``lifting" solutions has been relatively unexplored.
The original structural rounding paper~\cite{DEMAINE} mainly employs \naive lifting algorithms. For example, in \vc the entire edit set is added to the partial solution. We introduce a suite of new lifting algorithms for \vc and experimentally evaluate their quality, comparing them to one another as well as \naive and greedy approaches. We provide guarantees on the quality of these new lifting techniques relative to the optimal lift for a given partial solution.

Our experiments show that structural rounding outperforms traditional approximation techniques in a wide variety of graphs, and \naive and greedy lifting algorithms are competitive with more sophisticated methods.

We begin by discussing preliminaries and previous work, including the structural rounding framework and existing approaches for approximating \vc in Section~\ref{sec:prelims}.  We then describe the new lifting methods in Section~\ref{sec:algs}. We highlight interesting details of our implementation in Section~\ref{sec:implementation} and provide our experimental design and results in Section~\ref{sec:experiments}.
A complete version of this paper including information and tables in the appendix is available on the arXiv\footnote{https://arxiv.org/abs/1909.04611}.

%% file: sections/prelims.tex
We use $V$ and $E$ to represent the vertices and edges in a graph $G$, and let $n = |V|$ and $m = |E|$.
Much of this paper pertains to \vc, which asks for a set $S \subseteq V$ of minimum size such that every edge in $G$ has an endpoint in $S$.

\subsection{Structural Rounding}\label{sec:prelims-sr}

The structural rounding framework is simple: edit the input graph to a target class, apply an existing algorithm for that class (approximation or exact), and lift the partial solution to be a valid solution on the original graph. Under relatively broad assumptions, structural rounding extends algorithmic results for a structural class to graphs which are near that class.
The framework supports arbitrary graph edit
operations and both minimization and maximization problems, provided they
jointly satisfy two properties: a combinatorial property called ``stability''
and an algorithmic property called ``structural lifting''~\cite{DEMAINE}.
Roughly, these properties bound the amount each edit operation can change the optimal solution, but are parameterized to enable the
derivation of tighter bounds when the problem has additional structure.

Formally, structural rounding makes use of the following definitions; we restrict out statements to the minimization setting.

\begin{definition}\label{definition:gamma-close}
  A graph $G'$ is \emph{$\gamma$-editable} from a graph $G$
  under edit operation $\psi$
  if there is a sequence of $k \leq \gamma$ edits
  $\psi_1, \psi_2, \dots, \psi_k$ of type $\psi$ such that
  $G' = \kPsieditsG$.
  A graph $G$ is \emph{$\gamma$-close} to a graph class $\mC$
  under $\psi$ if some $G' \in \mC$ is
  $\gamma$-editable from $G$ under~$\psi$.

\end{definition}

We follow~\cite{DEMAINE} and use $\cost_{\Pi}$ to represent the cost function for a problem $\Pi$ and $\optsol{\Pi}{G}$ for the value of optimal solution to the problem $\Pi$ on graph $G$.

\begin{definition}
\label{definition:stable}
A graph minimization problem $\Pi$ is \emph{stable}
under an edit operation $\psi$ with constant $c'$
if $\optsol{\Pi}{G'} \leq \optsol{\Pi}{G} + c' \gamma$ for any graph $G'$ that is $\gamma$-editable from $G$ under~$\psi$.
\end{definition}

\begin{definition}
\label{definition:struct-lift-c-psi}
A minimization problem $\Pi$ can be \emph{structurally
lifted} with respect to an edit operation $\psi$ with constant $c$ if,
given any graph $G'$ that is $\gamma$-editable from $G$ under~$\psi$,
and given the corresponding edit sequence $\psi_1, \psi_2, \dots, \psi_k$
with $k \leq \gamma$,
a solution $S'$ for $G'$ can be converted in polynomial time to
a solution $S$ for $G$ such that
$\cost_{\Pi}(S) \leq \cost_{\Pi}(S') + c\cdot k$.
\end{definition}

We can now state the main result of structural rounding (Theorem 4.1) in~\cite{DEMAINE}, which pertains to \Editfullp, finding a smallest set of edits of type $\psi$ for editing to class $\mathcal{C}_\lambda$.

\begin{theorem}
\label{thm:general-edit-sr}
Let $\Pi$ be a minimization problem that is stable
under the edit operation $\psi$ with constant $c'$ and
that can be structurally lifted with respect to $\psi$ with constant~$c$.
If $\Pi$ has a polynomial-time $\rho(\lambda)$-approximation algorithm in the graph class $\mCp$,
and \Editfullp has a polynomial-time $(\alpha,\beta)$-approximation algorithm,
then there is a polynomial-time
$((1 + c' \alpha \delta) \cdot \rho(\beta \lambda) + c \alpha \delta)$-approximation
algorithm for $\Pi$
on any graph that is $(\delta \cdot \optsol{\Pi}{G})$-close
to the class~$\mCp$.
\end{theorem}

We note that an $(\alpha, \beta)$-approximation is one which uses at most $\alpha$ times the optimal number of edits needed to reach a graph in class $\mCp$ and returns a graph which is in class $\mC_{\beta \lambda}$ (i.e. if $\mCp$ is the class of graphs with treewidth at most $\lambda$, the returned graph has treewidth at most $\beta \lambda$). Demaine et al. demonstrated precisely how the ingredients of the framework fit together to obtain new approximations~\cite{DEMAINE}, and we refer the interested reader to this work for more details.

We consider \vc under the edit operation of vertex deletion. As noted in~\cite{DEMAINE}, the stability constant is $c'=0$ (deleting vertices cannot increase the size of a minimum size cove), and the lifting constant is $c=1$ for \vc, since the union of the deleted vertices and any solution $S'$ on $G'$ is a vertex cover of $G$.
Thus, by Theorem~\ref{thm:general-edit-sr}, \vc has a $(1 + \sqrt{\log(n)}\delta)$-approximation for any graph $G$ with an edit set of size $(\delta \optsol{\textsc{VC}}{G})$.

\subsection{\vc 2-Approximations}

\vc admits two straight-forward $2$-approximations. The best known, referred to here as \texttt{standard}, was discovered by both Gavril and Yannakakis~\cite{PAPADIMITROU}, and repeatedly selects an arbitrary edge $(u,v)$ in $G$, places both endpoints in the cover, and sets $G = G[V \setminus \{u, v\}]$. It is clear that whenever an edge is removed from $G$ it is incident to a node which has been added to the cover, thus every edge is covered.
Note that every edge must have at least one of its two endpoints in the cover.
Since every edge that we selected is independent from the other edges selected, any optimal solution must contain at least one of the two nodes from each selected edge.

The second $2$-approximation, which we refer to as \texttt{DFS}, follows from the work of Savage~\cite{SAVAGE} and constructs a cover $C$ by taking all of the non-leaf vertices from a depth-first search tree of the graph. Clearly this is a vertex cover as every edge touches an internal vertex. Further, the DFS tree has a matching with at least $|C|/2$ edges. Because these edges are independent, at least one endpoint must be a part of any valid cover, and thus $C$ is a $2$-approximation.

We compare the effectiveness of structural rounding with both of these $2$-approximations. There are also more complicated methods which yield slightly better approximation factors; namely a $2 - \Theta(1 / \sqrt{\log|V|})$-approximation~\cite{KARAKOSTAS} and a $2/(1 + \delta)$-approximation in $\delta$-dense graphs~\cite{KARPINSKI}. We do not compare against these algorithms due to their higher complexity.

\subsection{Odd Cycle Transversals}

An \emph{odd cycle transversal} is a set of vertices whose removal leaves a bipartite graph. Thus, graphs with ``small" odd cycle transversals, or \emph{OCT sets}, are referred to as ``near-bipartite," and this notion naturally arises in many applications~\cite{GULPINAR,PANCONESI,SCHROOK}.
We use $O$ to generally denote an OCT set, and given an OCT set for a graph $G=(V,E)$, we let $B = V \setminus O$. The subgraph induced on $B$ is bipartite with parts $L$ and $R$; and we let $n_L = |L|$, $n_R = |R|$, and $n_O = |O|$ for an OCT decomposition $(O,L,R)$.
While finding a smallest OCT set is NP-hard, efficient parameterized~\cite{IWATA, LOKSHTANOV} and approximation~\cite{AGARWAL} algorithms exist. We point out that the approximation factor in~\cite{AGARWAL} is $O(\sqrt{\log(n)})$, and thus using this as our editing algorithm in structural rounding leads to an approximation factor which is $\Omega(1)$.

While finding an optimal OCT set is hard, recent implementations~\cite{GOODRICH} of a heuristic ensemble alongside algorithms from ~\cite{AKIBA,HUFFNER} alleviate concerns that finding a good OCT decomposition creates a barrier to usability. For the sake of testing the efficacy of structural rounding, the OCT sets do not need to be optimal. We make use of a simple algorithm to find OCT sets: find two disjoint maximal independent sets and let the remaining vertices be our OCT set. See Section~\ref{sec:oct} for more details on how we carry out this process.

%% file: sections/algs.tex
In this section, we describe different lifting strategies for using structural rounding on \vc. To formalize notation, we let $\partialsol$ denote the solution on the edited-to instance $\editgraph$, and let $\editset$ be the edit set. We let the non-selected vertices from $\editgraph$ be denoted $\leftover$. Each lifting strategy finds a set $\liftsol$ such that $\liftsol \cup \partialsol$ is a solution on $G$.
Since we work with vertex deletions in this paper, $\editset \subseteq V(G)$ and $V(\editgraph) \cup \editset = V(G)$.
For a given $\partialsol$, we use $\optlift$ to denote the optimal lift, which is used to measure the quality of the various lifting strategies when appropriate.

\subsection{\Naive Lifting}

The simplest lifting strategy, \texttt{\naive} (from~\cite{DEMAINE}), is to let $\liftsol = \editset$. Clearly all edges in $G[\leftover \cup \editset]$ have at least one endpoint in $\editset$, thus this is a valid strategy. It also aligns with Theorem~\ref{thm:general-edit-sr} and Definition~\ref{definition:struct-lift-c-psi} giving lifting constant $c = 1$. However, there are no guarantees about the quality of this technique relative to $\optlift$. Depending on the structure of $\editset$, $|\optlift|$ may be $\Theta(|\editset|)$ or $\Theta(1)$, meaning that \texttt{\naive} could be near-optimal, or very far from it.

\subsection{Greedy Lifting}

A simple improvement upon \texttt{\naive} is to avoid inclusion of unnecessary vertices which already have all of their neighbors in the cover, which we refer to as the \texttt{greedy} technique.
We achieve this by  iterating over $\editset$ in an arbitrary order, maintaining a list of vertices outside the cover $U$, where initially $U = \leftover$ and $L = \emptyset$. For each vertex $v$, if it has a neighbor in $U$ we add it to $L$, otherwise we add it to $U$. Thus, in order for a vertex to be added to $L$, it must cover an edge where the other endpoint is not a part of the solution. Note that this strategy takes $O(m)$ time. While the returned solution is guaranteed to be no larger than the solution returned by \texttt{\naive}, it still suffers from not having any guarantee on solution quality relative to $\optlift$.

\subsection{2-Approximation Lifting}

While the two aforementioned strategies are good candidates for their simplicity and efficiency, we would prefer an approach with a theoretical guarantee. One natural candidate for lifting is using one of the \vc 2-approximations from Section~\ref{sec:prelims} on $G[\leftover \cup \editset]$, refered to as \texttt{apx}. Such a lift is guaranteed to be at most twice the size of $\optlift$. However, we note this may return solutions which are larger than $\editset$.

\subsection{\editfirst Lifting}

A natural extension to running a 2-approximation on $G[\leftover \cup \editset]$ is to instead run it on $G[\editset]$, obtaining a partial lift solution, $\liftsoltwo$. Then, if we let the nonselected vertices from $\editset$ be denoted as $\editset'$, $G[\editset' \cup \leftover]$ is a bipartite graph, and we can exactly solve in polynomial time~\cite{HOPCROFT} and find $\liftsolone$. Note that $\liftsoltwo \cup \liftsolone$ form a valid lifting solution. We are able to compute an approximation factor on the quality of this lift by comparing the sizes of $\liftsolone$ and $\liftsoltwo$.

Note that because each edge branched on in \texttt{standard} (or in the matching if using \texttt{DFS}) is independent, at least one vertex per edge must be included as part of any vertex cover. Thus, any lifting solution must have size at least $|\liftsoltwo|/2$. Further, all of these edges are independent from the edges in $G[\editset' \cup \leftover]$. When covering the edges in $G[\editset' \cup \leftover]$, no fewer than $|\liftsolone|$ vertices can be used. Thus, any valid lifting solution must contain at least $|\liftsoltwo|/2 + |\liftsolone|$ vertices and $|\optlift| \geq |\liftsoltwo|/2 + |\liftsolone|$.

If we let $p$ be the proportion of vertices added during Hopcroft-Karp ($p = |\liftsolone|/(|\liftsolone|+|\liftsoltwo|)$), we can rewrite our lower bound on $|\optlift|$ as  $(|\liftsolone| + |\liftsoltwo|)( (1-p)/2 + p) = (|\liftsolone| + |\liftsoltwo|) (1+p)/2$. Thus, $2|\optlift|/(1+p) \geq |\liftsolone| + |\liftsoltwo|$, and \editfirst lifting yields a $2/(1+p)$-approximation. Note that the approximation factor will range from $1$ to $2$.\looseness=-1

\subsection{\bipfirst Lifting}

A logical alternative to \editfirst lifting is to flip the order of the steps, and first solve on the bipartite subgraph induced on the edges between $\leftover$ and $\editset$ before running a $2$-approximation on the remaining edges. We refer to this method as \bipfirst lifting. Once again, we let the set of vertices which are given by the bipartite (exact) phase be denoted $\liftsolone$ and the vertices given by the approximation portion of the algorithm be denoted $\liftsoltwo$.

Unfortunately, we are unable to obtain an approximation factor as we did with \editfirst lifting. While it is clear that any lift must contain at least $\liftsolone$, we cannot make any stronger guarantee. As a simple example, if the edges between $\leftover$ and $\editset$ form an even-length cycle and every other vertex on the cycle has a pendant, Hopcroft-Karp may return the cycle vertices without pendants. Then, \bipfirst lifting would select all of the vertices, which is three times more than was necessary.

If we define $p$ as we did before ($p = |\liftsolone|/(|\liftsolone|+|\liftsoltwo|)$), we can still obtain an approximation factor better than $2$ when $p$ is large, namely $1/p$. Thus, when $p > 0.5$, we get better than a $2$-approximation. We speculate that this method will still prove useful in practice as there are likely to be many edges between $\leftover$ and $\editset$ when the edit set is found via any reasonable technique.

\subsection{Recursive Lifting}

One final type of lift comes from applying structural rounding as the lifting method. That is, after finding the partial solution in $\editgraph$, run structural rounding again. The precise manner in which this occurs can be done in several different ways. The first, \rec, is to completely re-run structural rounding on $G[\leftover \cup \editset]$ by finding a new editset on this subgraph and repeat. The second, \recedit, is to run structural rounding on $G[\editset]$, after which the only uncovered edges will form a bipartite graph between $\editset$ and $\leftover$ which can be solved exactly. A final option, \recbip, would be to first exactly solve on the edges between $\editset$ and $\leftover$ and then run structural rounding on any uncovered edges in $G[\editset]$. For each of these methods, we use \texttt{\naive} in the structural rounding component of the approach, but any of the aforementioned lifting techniques could be used. Further, while we only do one recursive step, one could recursively use structural rounding as the lift until exhaustion. It is worth noting that the approximation factor of these approaches depends on how well you are able to edit to the class of interest when rounding.

%% file: sections/setup.tex
In this section, we describe our datasets, along with how we measure approximation quality, and the hardware used for all experiments.
In order to experimentally evaluate the effectiveness of structural rounding, we executed each of the aforementioned \vc approximation
algorithms on a large corpus of synthetic ``near-bipartite" graphs with varied structure (controlled by generator parameters, see Section~\ref{sec:data}), as well as a collection of 130 real-world networks.

\subsection{Synthetic Data}\label{sec:data}

Recall that we use $n_O$, $n_L$, and $n_R$ to denote the sizes of $O$, $L$, and $R$ in an OCT decomposition where $O$ is the OCT set and $L$ and $R$ are the bipartite parts.
For convenience, throughout this section, we assume $n_L \geq n_R$ and let $n_B = n_L + n_R$. Our synthetic data was generated using a modified version of the random graph generator of Zhang et al.~\cite{ZHANG} that augments random bipartite graphs to have OCT sets of known size. The generator allows a user to specify the sizes of $L$, $R$, and $O$, the expected edge densities between $L$ and $R$, $O$ and $L \cup R$, and within $O$, and the coefficient of variation ($cv$; the standard deviation divided by the mean) of the expected number of neighbors in $L$ over $R$ and in $L \cup R$ over $O$. The generator is seeded for replicability.

Initially, we created graphs with each combination of $n_L/n_R \in \{1, 2, 10, 100\}$, expected edge density within $O \in \{0.001, 0.01, 0.05\}$, between $O$ and $L \cup R \in \{0.01, 0.05\}$, and between $L$ and $R \in \{0.001, 0.005, 0.01\}$, $n_O/n \in \{0.01, 0.05, 0.1, 0.25, 0.4\}$, and $cv$ of the expected number of neighbors in $L \cup R$ over $O \in \{0.5, 1.5\}$. We set $n_L, n_R,$ and $n_O$ based on the other parameters so that the expected value of $m$ was $4$ million.

We also generated graphs with 40 million and 200 million expected edges after paring down the parameter space to eliminate variants with little to no impact on solution quality.
We provide more details on the parameter down-selection in Section~\ref{sec:params}. On the larger graphs, we only ran the 2-approximations, \texttt{\naive} and \texttt{greedy} lifting, and the most effective lifts from the $4$ million edge data, \editfirst and \recedit. 
We also created a collection of synthetic graphs with 100K edges over the entire parameter space for the purpose of testing different implementations.

We further distinguish two scenarios in the 4M corpus.
While each graph has an OCT decomposition which corresponds to the specified parameters given by the generator, it does not mean that the OCT heuristic employed will find this partition.
Thus, we ran our algorithms on the $4$ million edge graphs with two approaches; (i) with the OCT decomposition \emph{prescribed} to be the exact one given by the generator (4M-pre) and (ii) with an OCT decomposition which is \emph{procured} by our heuristic (4M-pro). We observed that differences in outcomes were minor (see Figure~\ref{fig:all-data}) and use only approach (ii) on the 40M and 200M corpuses because of it being a more realistic scenario in practice.
The prescribed approach does have the benefit of allowing us to determine the impact that the different generator parameters had on the solution quality of each strategy, which we observe in Section~\ref{sec:params}.

\subsection{Real World Data}

\begin{table}
{\small
\begin{center}
    \begin{tabularx}{0.5\textwidth}{@{\extracolsep{\fill}} l|cccc}
       & $n$ & $m$ & avg. degree & $|O|/n$\\
	\midrule
	min & 9	& 8 & 1.77 & 0\\
	max & 556686 & 987091 & 326.851 & 0.984\\
	median & 1104.5 & 2984 & 4.867 & 0.256\\
	mean & 18859.164 & 64881.578 & 14.092 & 0.343\\
        \bottomrule
    \end{tabularx}
\end{center}
}
    \caption{\label{tab:rw}
Summary statistics of real-world corpus.
    }
\end{table}

The real world corpus is a set of 130 graphs from various domains including social networks, physical infrastructure, biology, diseases, and compilation dependencies taken from~\cite{GRAPHS}.
These networks vary widely in size, average degree, and relative OCT size (see Table~\ref{tab:rw}).  We note that despite the variation in average degree, most of the graphs in the real world corpus are substantially sparser than any of the synthetic graphs used.

\subsection{Measuring Quality}
In order to measure the quality of an approximate solution, we compute the ratio of its size to half of the ``worst-case" 2-approximation (the maximum cover size found by any 2-approximation run); we refer to this as the \emph{approximation ratio}.
For example, if some 2-approximation returns a cover of size 100, we know an optimal solution has at least 50 vertices, and a cover of size 75 is at-worst a 1.5-approximation.

\begin{table}[b]
{\small
\begin{center}
	\hspace{1.7cm}\textbf{Method}
    \begin{tabularx}{0.5\textwidth}{@{\extracolsep{\fill}} lcccc}
        & \texttt{std-high} & \texttt{std-low} & \texttt{std-rand} & \texttt{DFS}\\
	\midrule
	avg. size & 3340.617 & 3647.758 & 3531.626 & 3397.374\\
        \bottomrule
    \end{tabularx}
\end{center}
}
\caption{\label{tab:2-apx}
Mean approximation sizes for 2-approximation variants on the 100K corpus. We employ three edge selection techniques for \texttt{standard}: incident to a high-degree vertex (\texttt{std-high}), incident to a low-degree vertex (\texttt{std-low}), and arbitrarily (\texttt{std-rand}).
}

\end{table}

\subsection{Hardware}
All experiments were run on identical hardware; each server had four Intel Xeon E5-2623 v3 CPUs (3.00GHz)
and 64GB DDR4 memory. The servers ran Fedora 27 with Linux kernel 4.16.7-200.fc27.x86\_64.
The code is entirely written in Python 3 and run using version 3.6.5.
Our code is open source under a BSD 3-clause license and publicly available~\cite{LAVALLEE}.

%% file: sections/implementation.tex
\begin{figure}
	\includegraphics[width=0.5\textwidth]{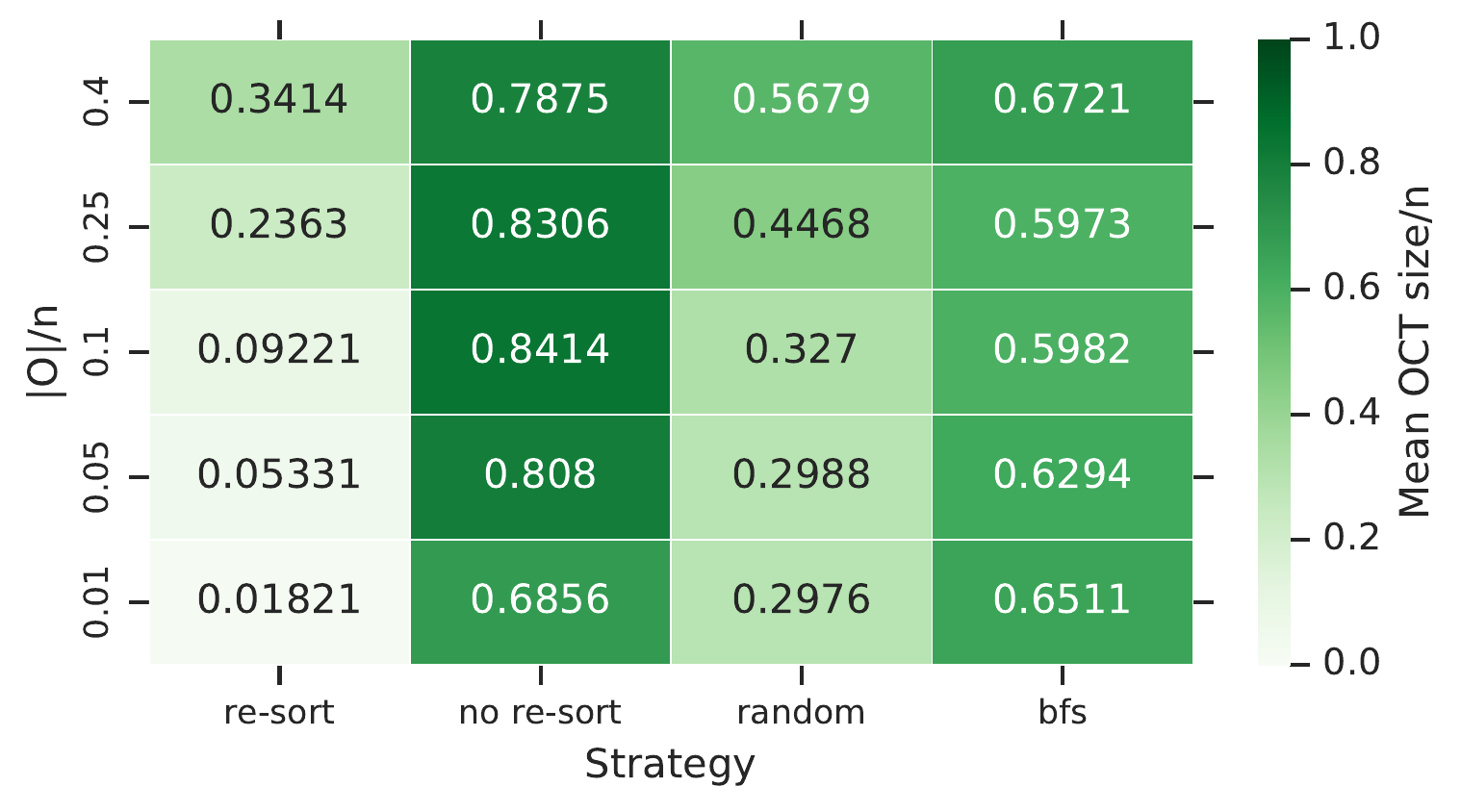}%
  \caption{\label{fig:octsize-100}
	Mean relative OCT sizes for OCT decomposition algorithms on the 100K corpus, partitioned by prescribed OCT ratio.
	Approaches include a BFS-based technique (bfs), and the construction of two maximal independent sets with three vertex selection rules: arbitrarily (random), by minimum initial degree (no re-sort), and by minimum degree with re-sorting (re-sort).
  }
\end{figure}

\begin{figure}
	\includegraphics[trim=0 1em 0 3em,clip,width=0.5\textwidth]{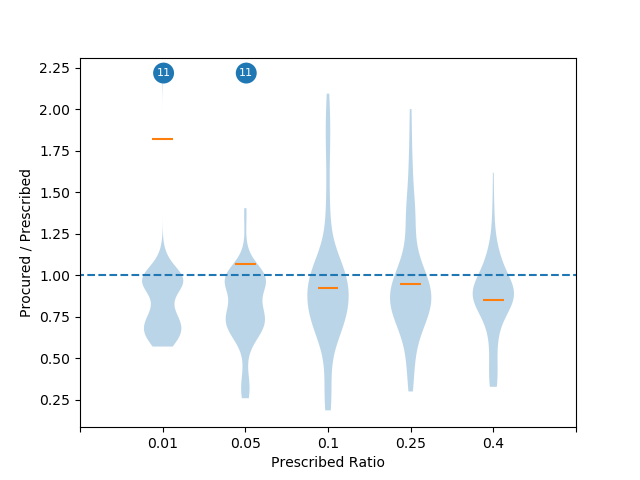}
	\caption{Distribution of procured OCT sizes relative to prescribed OCT sizes found on the 100K corpus via the heuristic OCT algorithm with re-sorting. The mean of each distribution is given by the orange bars on the plot. 11 outliers were collected and shown at 2.25 in the first two distributions to preserve the scale. Note that these values are still considered in the mean.  A ratio of 1 or smaller indicates that the heuristic OCT set approach found a better OCT set than the one prescribed by the generator.}
	\label{fig:right-octsize}
\end{figure}

\begin{figure*}
	\begin{minipage}{0.5\textwidth}
		\begin{centering}
			\includegraphics [width=\textwidth]{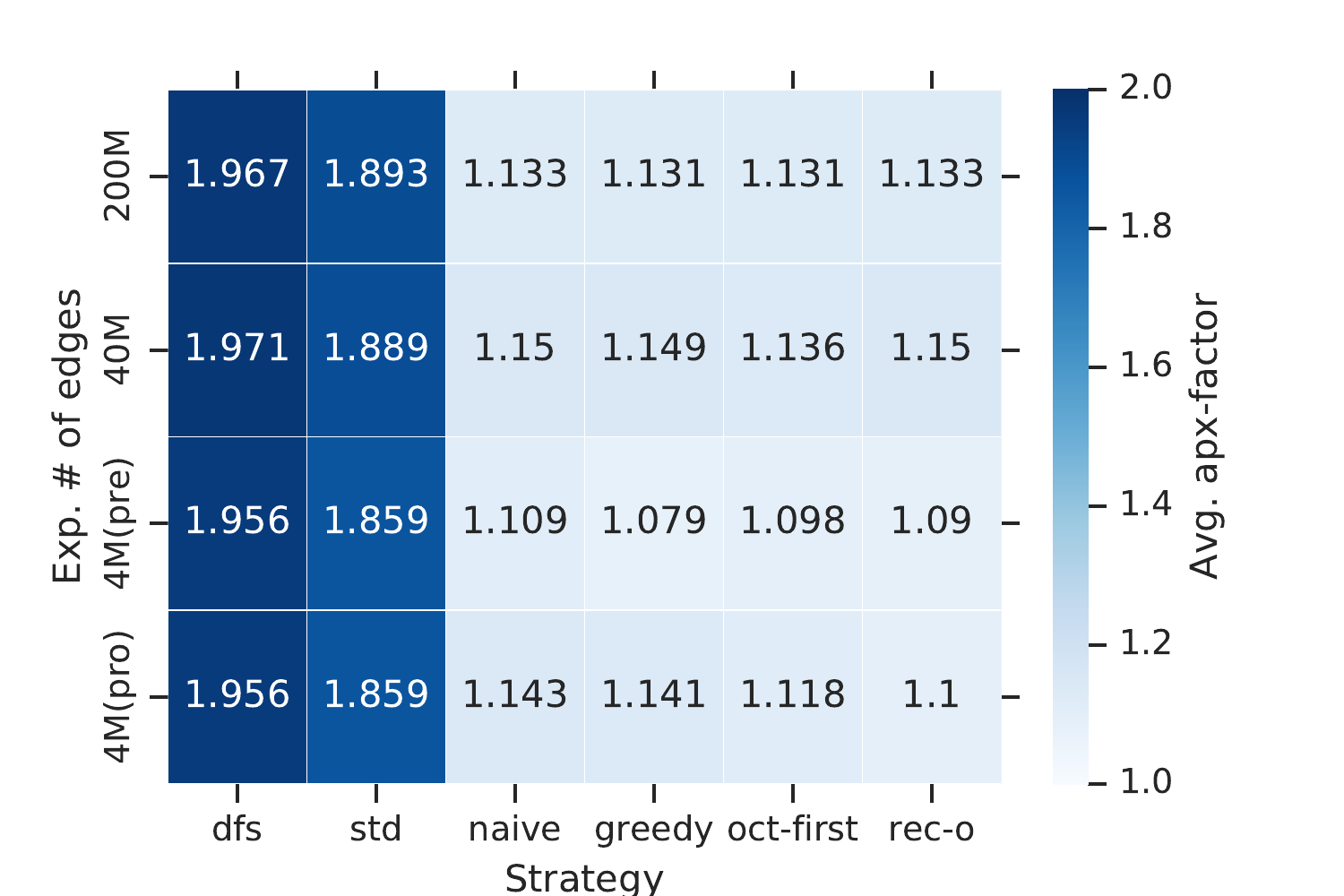}%
		\end{centering}
	\end{minipage}
	\begin{minipage}{0.5\textwidth}
		\begin{centering}
			\includegraphics[width=\textwidth]{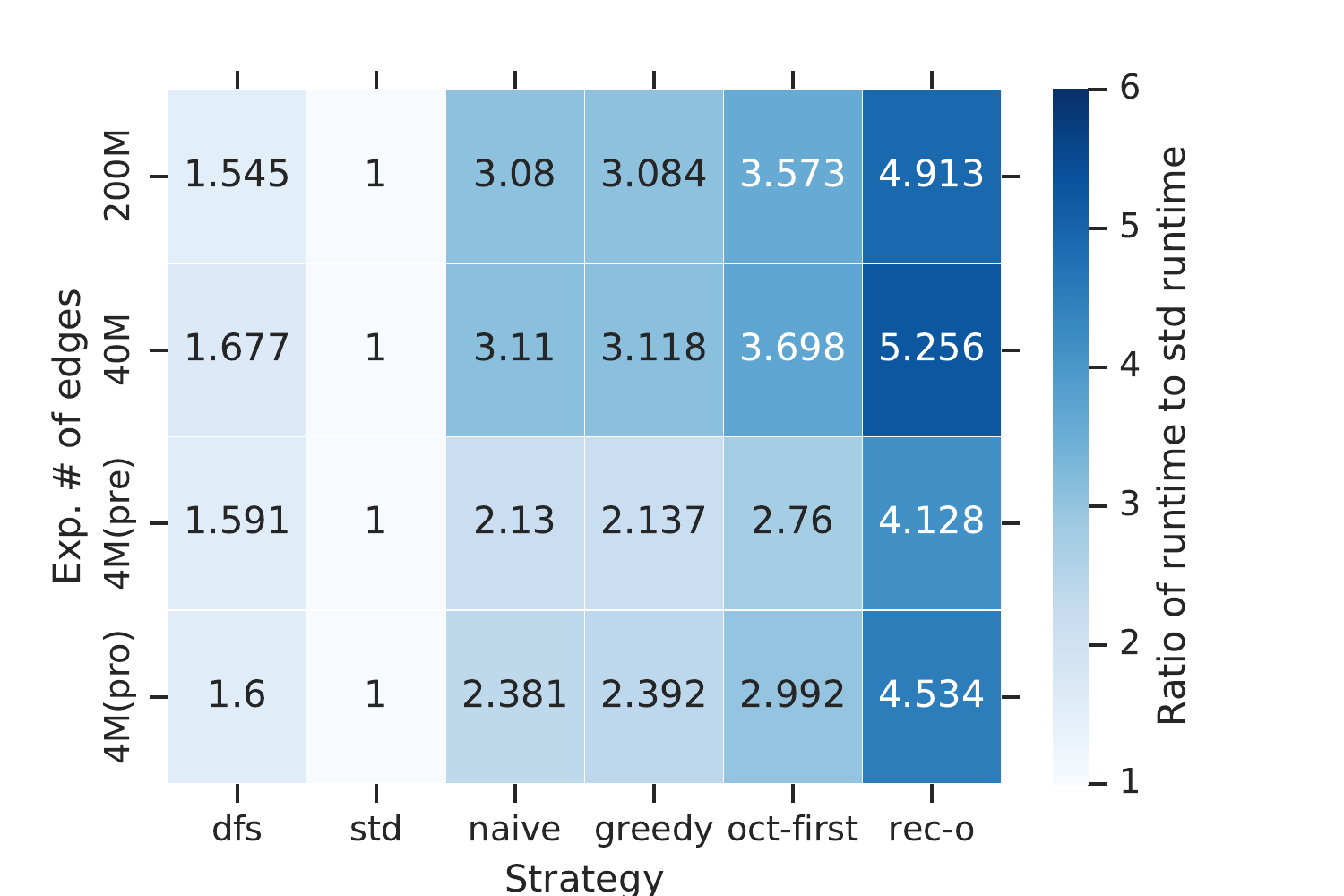}
		\end{centering}
	\end{minipage}

	\caption{\label{fig:all-data}
	Average approximation ratios (left) and relative runtime (right) on the full synthetic corpus, partitioned by expected number of edges. For 4M edge graphs, we use both procured (``pro") and prescribed (``pre") OCT decompositions. Runtimes are given relative to that of \texttt{standard}, which is fastest on average (mean runtimes: 1.511s on 4M-pro, 1.502s on 4M-pre, 20.551s on 40M, and 136.779s on 200M). For structural rounding approaches, the time it takes to find the OCT decomposition \emph{is} accounted for in the runtime (note that for 4M-pre, this contributes zero since the decomposition is given as input).
	}
\end{figure*}

In this section, we provide details on algorithm implementation along with data supporting experimental design decisions.

\subsection{2-Approximation Variants}\label{sec:standard}

We tested four different 2-approximations for \vc including \texttt{DFS} of~\cite{SAVAGE} and three variants of \texttt{standard} over all of the synthetic graphs with 100K edges in expectation.
Specifically, the three \texttt{standard} variants differed on how edges were selected.
The edge selection techniques tested were random selection, a random edge incident on a high degree vertex, and a random edge incident on a low degree vertex which we will refer to as \texttt{std-rand}, \texttt{std-high}, and \texttt{std-low} respectively.
As seen in Table~\ref{tab:2-apx}, the best performing 2-approximation for \vc was generally to use \texttt{std-high}.
This approach almost always outperformed \texttt{std-rand} on the synthetic graphs.
Conversely, \texttt{std-low} almost always performs worse, providing us with the highest possible lower bounds on the optimal solution size.
Using \texttt{DFS} did occasionally produce smaller solutions than \texttt{standard}.
However, in over 50\% of parameter settings it performed worse, and likewise on average, it produced larger solutions.
Since \texttt{DFS} is also randomized, we chose to use \texttt{standard} to maximize performance and not require additional trials for each graph.

\subsection{Finding OCT Sets}\label{sec:oct}

We compared two heuristic approaches to finding small OCT sets on the 100K corpus.
We found that greedily constructing two maximal independent sets generally outperformed using a BFS-search, where we fix a BFS-ordering, and greedily add each vertex in order to the left or right if possible, and otherwise to the OCT set (see Figure~\ref{fig:octsize-100}).
We note that the performance of the maximal independent set heuristic crucially depends on not only choosing minimum degree vertices first, but also on updating the degree of each vertex and re-sorting as vertices are added to the independent set.
When the vertices are not re-sorted or are chosen randomly, we frequently found OCT sets that far exceeded what we expected based on the generator's parameter settings.
When testing the BFS-coloring heuristic, we notice that despite offering a speed advantage over the re-sorting algorithm, the OCT sets produced are almost always substantially larger.
Furthermore, BFS-search loses its speed advantage when random vertices are chosen while still producing larger OCT sets.
Since heuristically finding an OCT set is rarely the bottleneck in a structural rounding approximation, we opted to use the maximal independent set strategy with re-sorting.

Even when we use the maximal independent set heuristic with re-sorting, we have no guarantees that the algorithm will produce the same OCT set as the one prescribed by the generator.
In Figure~\ref{fig:right-octsize} though, we show that in the majority of graphs in the 100K corpus, the heuristic OCT algorithm finds OCT sets of comparable size to those prescribed by the generator.

\subsection{Converting Maximum Matchings}

Unlike the popular \texttt{NetworkX} package in Python, our implementation for converting a maximal matching found in a bipartite graph to a vertex cover is done iteratively and not recursively, allowing it to be somewhat faster. We avoid using recursion when possible in all functions in our codebase, particularly in implementations of depth-first search.

\subsection{Variance Between Runs}~\label{sec:var}

We ran each algorithmic component of our framework 50 times on the 100K synthetic corpus to test the amount of variance in the runtimes. In all cases, the maximum ratio of variance to mean averaged over the corpus was at most 0.0004, giving us confidence that a single trial of each graph-algorithm combination would result in a representative runtime. We include all of the variance results in the appendix.

%% file: sections/experiments.tex
We begin by describing the three key findings from our empirical evaluation of the structural rounding framework, then discuss in greater detail the impact of structural variation in the input (as prescribed by generator parameters, Section~\ref{sec:params}), results in the procured decomposition setting (Section~\ref{sec:procured}), and finally results on the real-world corpus (Section~\ref{sec:rw}).

1. \emph{Structural rounding consistently outperforms traditional approximation techniques.} Both \texttt{standard} and \texttt{DFS} perform considerably worse than structural rounding-based methods when averaged over our entire corpus of graphs in both the prescribed and procured settings (left panel of Figure~\ref{fig:all-data}). Because structural rounding is a more involved technique, it tends to take at least twice as long to run compared to the $2$-approximations (right panel of Figure~\ref{fig:all-data}), but absolute runtimes remain very low and usable in practical settings.
The bottleneck for this configuration of structural rounding is solving vertex cover exactly in the edited graph which can take up to 30 seconds in the 200M corpus.
Both \texttt{\naive} and \texttt{greedy} lifting finish almost instantaneously even in the 200M corpus.

2. \emph{\texttt{\Naive} and \texttt{greedy} are often comparable with more sophisticated lifting approaches.} The quality of the simpler lifts for \vc are roughly the same (and sometimes better) than that of the other structural rounding lifts (left panel of Figure~\ref{fig:all-data}). Given the additional time (right panel of Figure~\ref{fig:all-data}) and memory the more advanced lifts require, \texttt{greedy} combines the efficiency of traditional techniques and the accuracy of more sophisticated lifting.

\begin{figure}
\includegraphics[trim=5em 0 0 0,clip,width=0.575\textwidth]{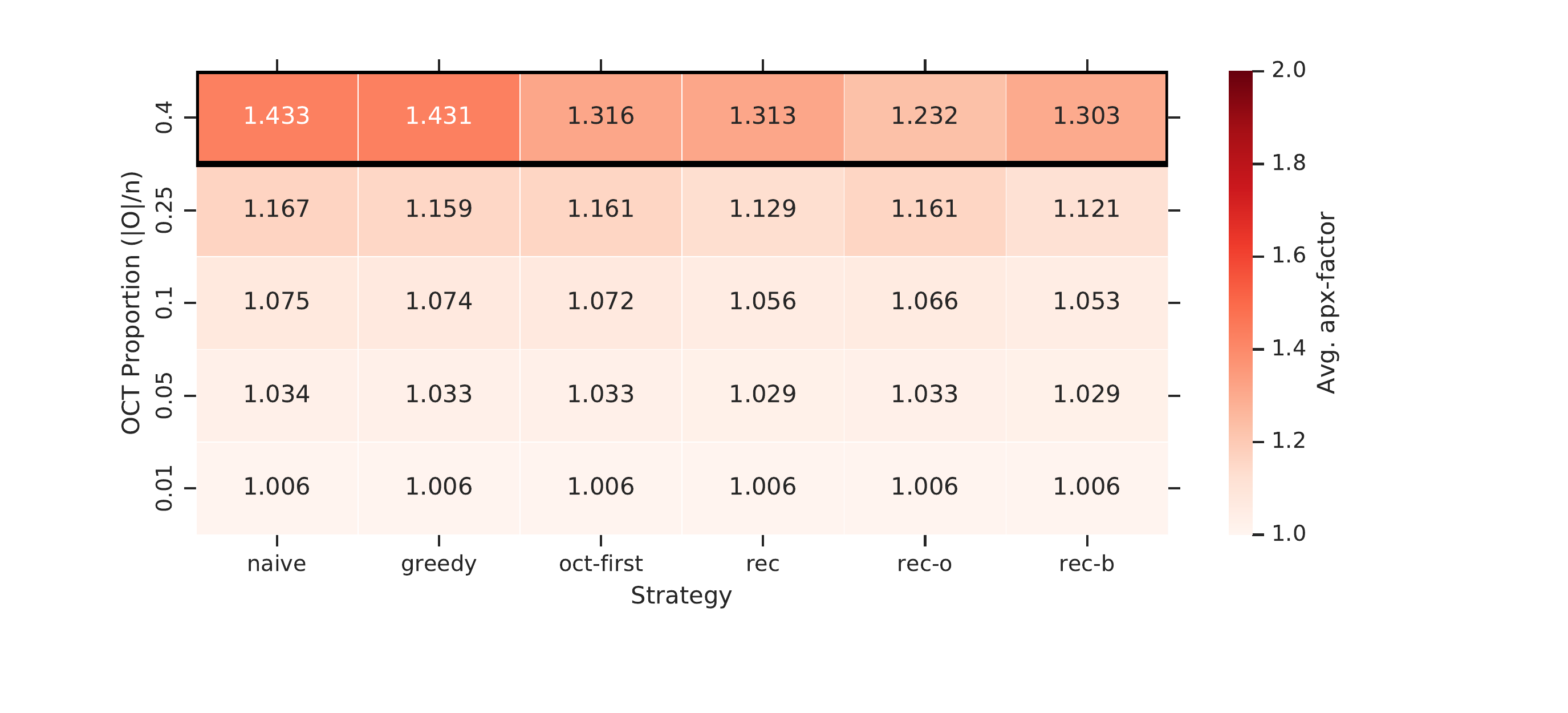}
\includegraphics [trim=5em 0 0 0,clip,width=0.575\textwidth]{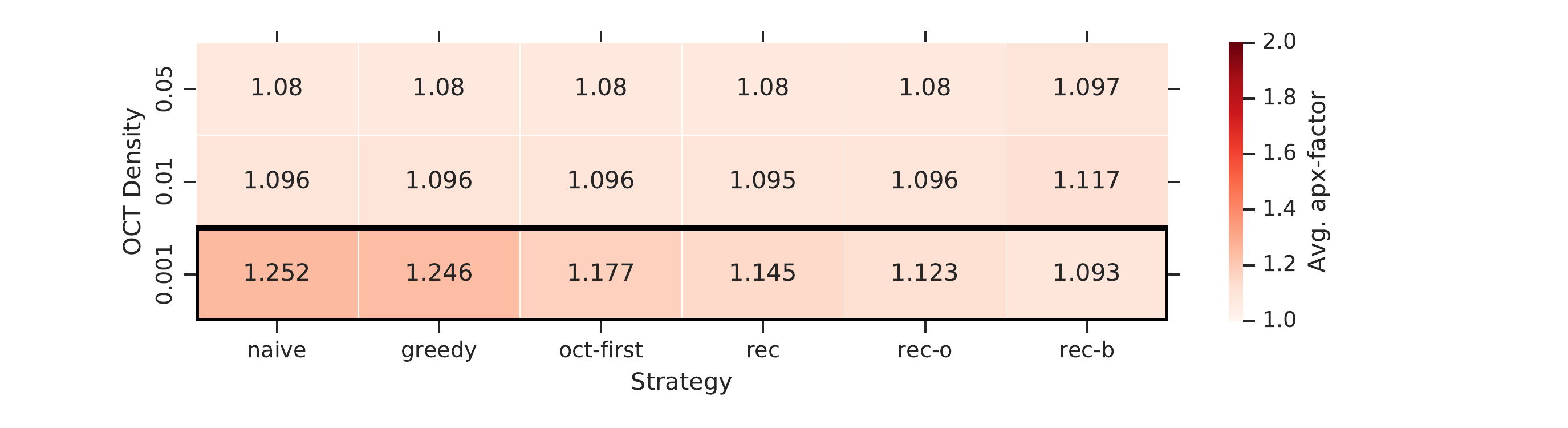}
    \caption{\label{fig:smarter}
Two settings when more sophisticated lifting strategies outperform \texttt{\naive}/\texttt{greedy} on the 4M-pro data.
We partition graphs based on two generator parameters: relative OCT set size (top), and expected edge density within $O$ (bottom).\looseness-1 }
\end{figure}

3. \emph{There are settings where the more sophisticated lifting approaches outperform \texttt{\naive} and \texttt{greedy}.} In the procured setting, the smarter lifting techniques outperform \texttt{\naive} and \texttt{greedy} by the most when the ratio $n_L/n_R$ is largest, the $cv$ between $O$ and $\{L,R\}$ is largest, the expected edge density in $O$ is smallest, and the the proportion of the graph in $O$ is largest (see Figure~\ref{fig:smarter} for examples of the latter two). This behavior is seen on a lesser scale on the larger graphs. In the prescribed setting, the difference is less pronounced and \recbip tends to distinguish itself the most from \texttt{\naive} and \texttt{greedy}, namely when the ratio $n_L/n_R$ or the $cv$ between $O$ and $\{L,R\}$ is largest, or the proportion of the graph in $O$ is smallest.

We include all of the results from our experiments in the appendix. Throughout the paper we color our figures based on the dataset they represent as follows; blue is an aggregate view of the large data ($\geq$ 4M expected edges), green is the supplementary data (100K expected edges), red is the procured setting of the 4M expected edge data (4M-pro), purple is the prescribed setting of the 4M expected edge data (4M-pre), orange is the 40M expected edge data, and grey is the 200M expected edge data.

\subsection{Impact of Generator Parameters}~\label{sec:params}

We now discuss how each of the generator parameters impact the various lifting methods. We utilize the prescribed data (4M-pre) to measure these effects, as this allows us to isolate the variables.

\textbf{$|$\emph{L}$|$/$|$\emph{R}$|$.} When this ratio is 2, 10, or 100, \recbip lifting is the optimal strategy. However, when it is 1, it does significantly worse, and \texttt{greedy} is the best option. It is likely that this is because $I$, the set of vertices not selected when solving on the edited-to graph $G[L \cup R]$, is smaller than in the other cases. Then when \recbip solves on the edges between $X$ and $I$ it picks up most of $I$, which is unnecessary because most of these edges will be covered by the solution on $X$.
When compared to the worst-case 2-approximation, all of the structural rounding methods do best when the ratio is 2 and 10. Due to the lack of distinction between these two settings, we only use ratios of 1, 10, and 100 when generating larger graphs. \texttt{DFS} monotonically worsens as this ratio increases, and is always worse than \texttt{standard}.

\textbf{$cv$ between $O$ and $\{L,R\}$.} \recbip is the best method in the higher setting, while \texttt{greedy} is best in the lower setting. In general, the approximation factors of the structural rounding methods are lower when the $cv$ is lower. Interestingly, there is no difference between the settings when using 2-approximation lifting. Both 2-approximations also do better when the $cv$ is smaller.

\begin{figure}
\includegraphics [trim=5em 0 0 0,clip,width=0.575\textwidth]{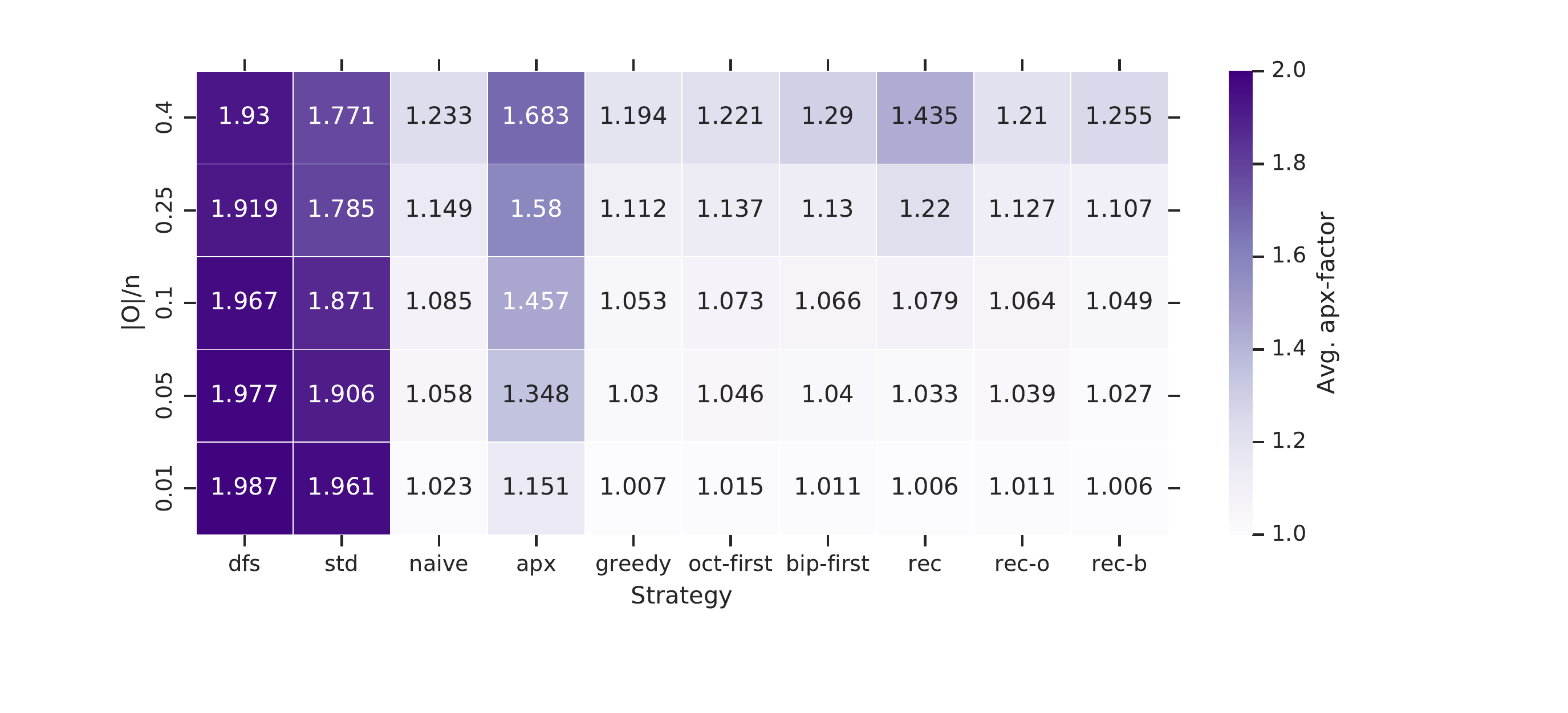}
\includegraphics [trim=5em 0 0 0,clip,width=0.575\textwidth]{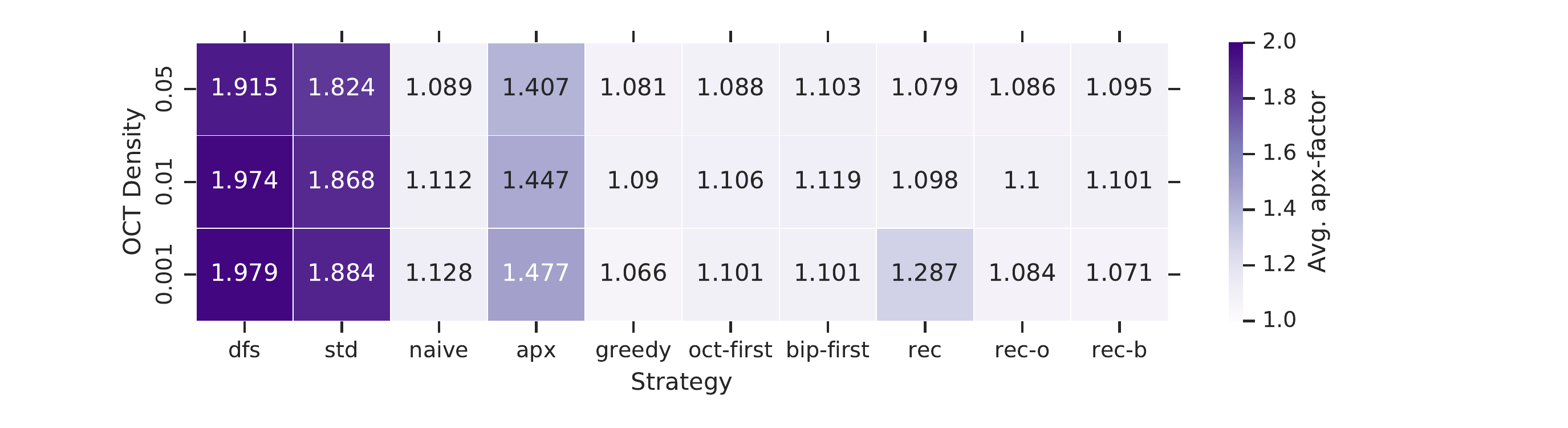}
    \caption{\label{fig:params}
Impact of the generator parameters on the various lifting methods. The top heatmap shows the effect of varying $|O|/n$ and the bottom heatmap shows the effect of varying the expected edge density within $O$.\looseness=-1
    }
\end{figure}

\textbf{$|$\emph{O}$|$/\emph{n}.} For the four smallest settings (see Figure~\ref{fig:params}, top), \recbip is among the best approximations. However, for the largest OCT ratio (0.4), \texttt{greedy} is the optimal approach, and \texttt{\naive}, \editfirst, and \recedit are all more effective than \recbip. This shift is likely due to the edit set $X$ being larger relative to $I$, the set of vertices not selected when solving on the edited-to graph $G[L \cup R]$. Similar to our observations when varying $|L|/|R|$, it is likely that \recbip unnecessarily adds a majority of $I$, inflating its solution size.

\textbf{Expected edge density in $O$.} \rec is the best strategy in the highest density setting, while \texttt{greedy} does best in the two lower settings (see Figure~\ref{fig:params}, bottom). In fact, \rec is by far the worst structural rounding approach for the lowest OCT density setting. This is likely due to the extreme sparsity in the edit set in this setting, and thus \rec finds a large OCT set when re-running structural rounding on $G[I \cup X]$, which in turn gets completely added to the solution due our using \texttt{\naive} for this step. One particularly interesting observation is that the structural rounding approaches tend to be split on whether they are best at higher or lower density settings, and five of them are worst on the moderate setting. This behavior was only observed for \recbip in the procured setting, and due to the minor difference between 0.01 and 0.05, we only used the two extreme settings when generating larger graphs. Both 2-approximations do best when the density is higher.

\begin{figure*}
	\begin{minipage}{0.5\textwidth}
		\begin{centering}
			\includegraphics[width=0.99\textwidth]{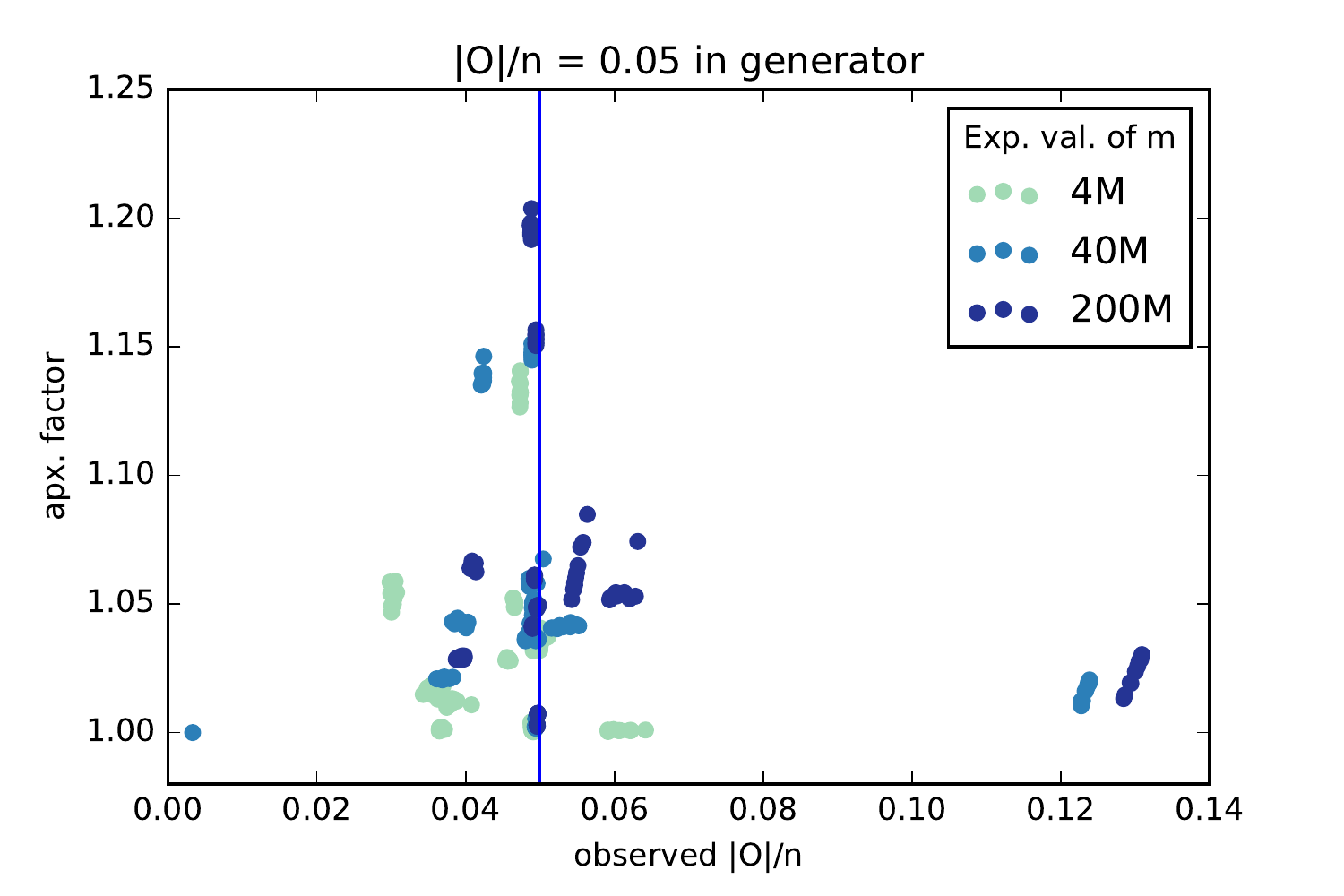}
		\end{centering}
	\end{minipage}
	\begin{minipage}{0.5\textwidth}
		\begin{centering}
			\includegraphics[width=0.99\textwidth]{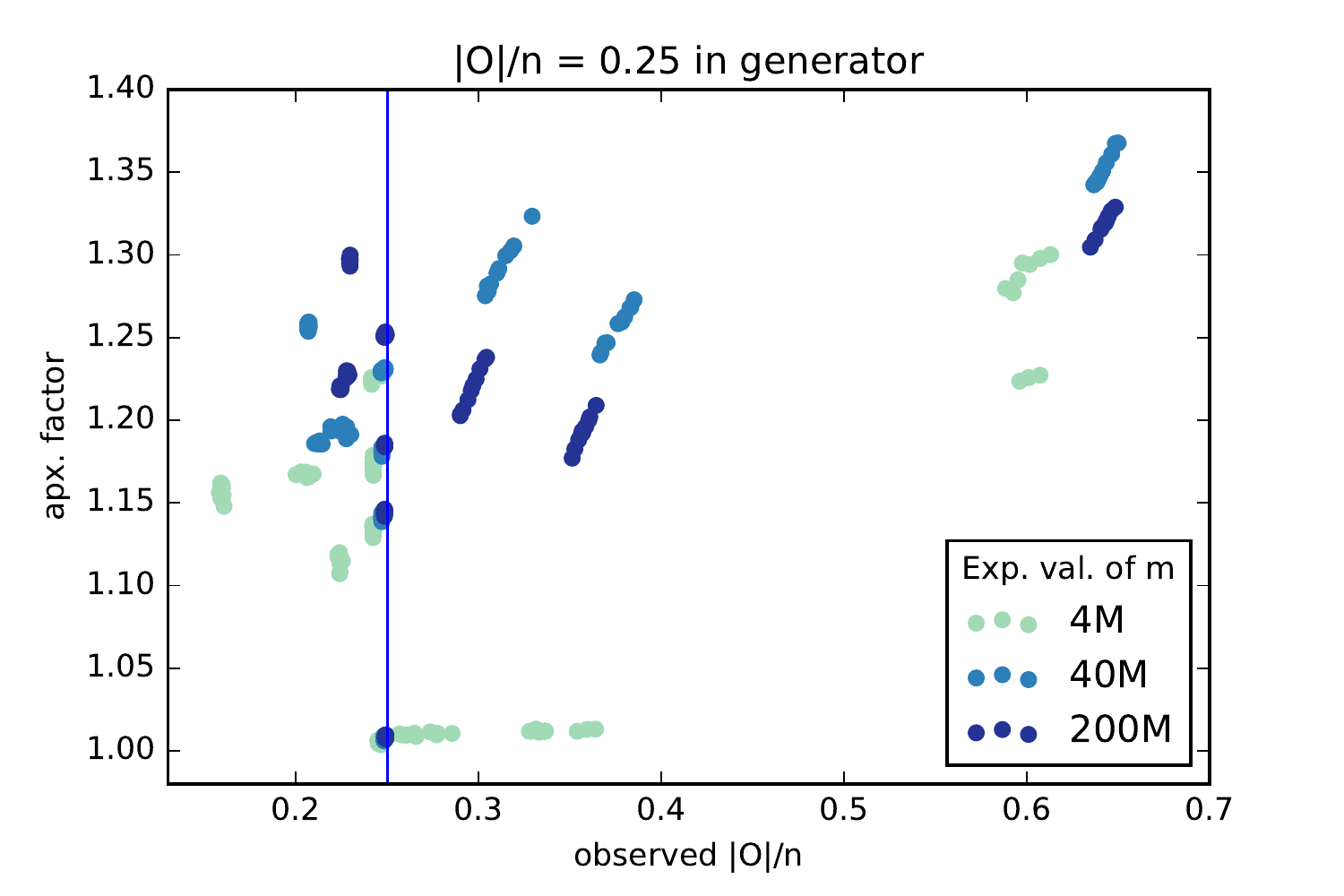}
		\end{centering}
	\end{minipage}

	\caption{\label{fig:octs}
		Correspondence between the size of the procured OCT set and solution quality across 4M-pro (green), 40M (blue), and 200M (red) corpuses; only graphs generated with parameters used in all experiments are included for consistency. The data is partitioned by prescribed relative OCT size $|O|/n$, with $0.05$ shown at left and $0.25$ at right. We observe
    that performance is most consistent when the procured decomposition has OCT size close to the prescribed value.
	}
\end{figure*}

\textbf{Expected edge density between $O$ and $\{L, R\}$.} In both settings, \texttt{greedy} is the best lifting strategy. All of the structural rounding approaches have better approximation factors when this expected density is smaller, as does \texttt{DFS}. However, \texttt{standard} does better in the larger density setting. The magnitude of these differences tends to be small, so we did not vary this value on the larger graphs.

\textbf{Expected edge density between $L$ and $R$.} Once again, \texttt{greedy} is the optimal strategy in all settings. There is minimal impact of the quality of structural rounding when compared to the worst-case 2-approximation, and thus we did not vary this parameter on the larger graphs. The 2-approximations also did not vary much based on this density.

\subsection{Procured Results}\label{sec:procured}

While our findings on the procured data are generally consistent with the prescribed data, we observe the following anomalies. When $n_L/n_R=2$, \recedit does slightly better than \recbip on the 4M edge graphs, while on the larger graphs \editfirst is the optimal approach for the larger ratio settings.  When the expected number of edges is 40M or 200M, \texttt{DFS} does best when $n_L/n_R=10$.
\rec is the optimal strategy when the $cv$ between $O$ and $\{L,R\}$ is $0.5$ for the graphs with 4M edges, while \editfirst is preferred in this setting on the larger graphs.
When $n_O/n=0.4$, the best approach on the 4M edge graphs is via \recedit, while \editfirst and \texttt{greedy} perform best on the 40M and 200M edge graphs under all settings. \rec is not the worst-performing method on the 4M graphs when the expected density with OCT is 0.0001, which is likely due to the observed OCT set not having this extreme sparsity. There is little difference over the OCT density among the structural rounding approximations on the 200M graphs, while \editfirst is best on the 40M graphs when the OCT density is low.
\recedit is the best strategy on the 4M edge graphs when expected density between $O$ and $\{L,R\}$ is higher, while \rec is optimal for the lower setting.
Regarding the expected edge density between $L$ and $R$, \recedit is the best lifting technique under all settings.

\begin{figure*}
	\begin{minipage}{0.5\textwidth}
		\begin{centering}
			\includegraphics[width=\textwidth]{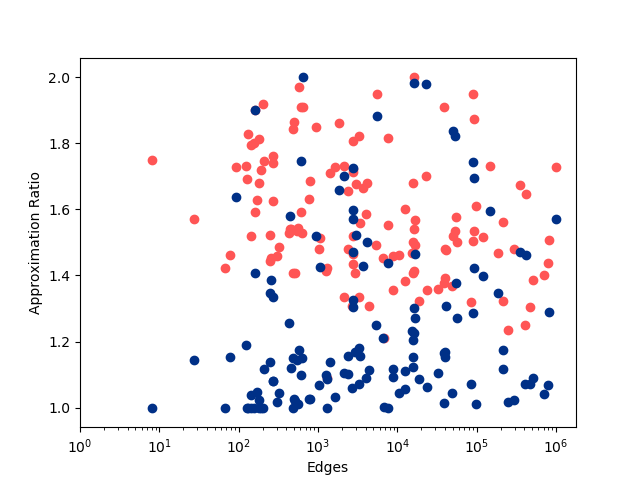}
		\end{centering}
	\end{minipage}
	\begin{minipage}{0.5\textwidth}
		\begin{centering}
			\includegraphics[width=\textwidth]{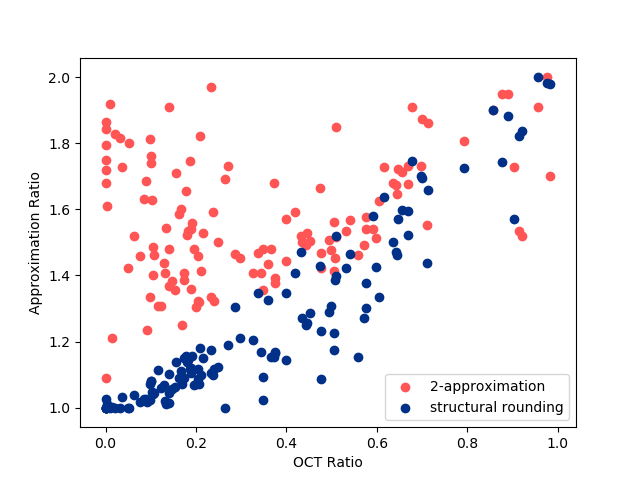}
		\end{centering}
	\end{minipage}

	\caption{\label{fig:real}
		The best approximation ratios achieved by 2-approximations (in red) versus structural rounding (in blue) on graphs from the real-world corpus.
		The ratios are computed by comparing the best 2-approximation or structural rounding result with the worst 2-approximation.
		The 2-approximations considered are \texttt{DFS}, std-high, and std-low.
		The lifting algorithms used are \texttt{greedy}, \editfirst, and \bipfirst.
		On the left, we plot these ratios against the number of edges in the graph on a log scale, while on the right, we plot the ratios against the proportion of the graph in the OCT set.
	}
\end{figure*}

We observe similar behavior at all three scales of the procured data when we distinguish the graphs based on specific generator parameter values. This is highlighted in the data in the appendix where we split the data based on the generator value of $n_O/n$. Generally the graphs of all three sizes behave comparably when the observed size of $O$ is close to that of the generator. When the procured proportion is greater than the prescribed value, we see more distinction among the sizes, as generally the found OCT sets are larger in the 40M and 200M settings (in all but the $n_O/n=0.4$ setting), resulting in slightly worse solution quality in the larger graphs.

\subsection{Real-world Experiments}~\label{sec:rw}

Lastly, we ran our collection of algorithms on a corpus of real-world graphs compiled from various scientific, sociological, and technological domains.
Since these graphs are much sparser than any of the synthetic graphs we tested (average degree less than 2 in some cases), finding the OCT set became a bottleneck in a few of the larger graphs.
To remedy this, we used random edge selection in the heuristic OCT algorithm to keep runtimes low.
Additionally, since the graphs are so sparse, this choice had minimal impact on the performance of the structural rounding algorithms.
Overall, we see that structural rounding outperforms the 2-approximations.

In Figure~\ref{fig:real}, we observe that structural rounding performs better in a large number of graphs across all sizes.
Furthermore, we see that when graphs have small OCT sets, structural rounding consistently produces smaller solutions.
Even on graphs with large OCT sets, structural rounding matches the solution quality of the 2-approximations, delivering well beyond its comparatively weak theoretical guarantees.
Lastly, we note that the structural rounding algorithms finish reasonably quickly, taking less than 10 seconds to run on graphs with nearly a million edges and over half a million vertices.

\subsection{Heuristic Comparison}~\label{sec:heur}

Although structural rounding outperforms common approaches for \emph{approximating} vertex cover, we also questioned whether structural rounding would be competitive against state of the art heuristics. In order to compare the performance of structural rounding against heuristics, we tested performance against a simple greedy \texttt{heuristic} which repeatedly adds the vertex covering the most uncovered edges until every edge is covered. We ran \texttt{heuristic} on the 4M-procured data set. In Table~\ref{tab:heur}, we see that \texttt{heuristic} ran in less time for graphs with small oct sets while producing solutions nearly as small as structural rounding. In graphs with larger oct sets, \texttt{heuristic} becomes slower than \texttt{greedy} and \texttt{\naive} lifting, but performs much better while still running faster than more sophisticated lifting techniques needed to achieve the best structural rounding results. In summary, for use cases where an approximation guarantee is not required, structural rounding is unlikely to be preferred over simple heuristics due to the increased runtime. The tradeoff with more sophisticated (and thus costly) heuristics is less clear and requires further experimentation.

%% file: sections/conclusion.tex
In this work we provide the first implementation and empirical evaluation of structural rounding
and demonstrate that structural rounding provides better approximations for \vc than traditional approaches on a large corpus of widely varying graphs.
These results highlight the importance of experimentally evaluating theoretical approaches. While the structural rounding framework for \vc under vertex deletion to bipartite graphs has weaker theoretical guarantees than the traditional methods, its practical performance is much stronger.
We introduced a suite of more sophisticated lifting algorithms but showed that the simple greedy approach was extremely competitive in most scenarios

While this paper concerns itself with editing to bipartite graphs and solving \vc, there is an abundance of interesting and unexplored directions. One might consider editing to graph classes other than bipartite, e.g. classes suggested in~\cite{DEMAINE} including bounded degeneracy and bounded treewidth.
The promise of more sophisticated lifting approaches on other problems such as \textsc{Independent Set}, \textsc{Dominating Set}, and \textsc{Feedback Vertex Set} is also unclear.
While we are interested in finding new lifting algorithms for problems other than \vc, the success of \texttt{\naive} and \texttt{greedy} indicated in this paper leads us to believe that it is reasonable to anticipate similar results for other problems. Finally, we would be interested in comparing the effectiveness of structural rounding when using other edit types, namely edge deletion and edge contraction, with that of using vertex deletions.

\clearpage

\begin{table}
{\small
\begin{center}
    \hspace{2.5cm}\textbf{OCT Ratio}
    \begin{tabularx}{0.5\textwidth}{@{\extracolsep{\fill}} l|ccccc}
       & .01 & .05 & .1 & .25 & .4\\
	\midrule
	\color{red}{\texttt{heuristic} time} & \color{red}{3.37}	& \color{red}{3.19} & \color{red}{3.09} & \color{red}{2.93} & \color{red}{2.90}\\
	\color{blue}{sr-\texttt{greedy} time} & \color{blue}{7.51} & \color{blue}{4.58} & \color{blue}{3.37} & \color{blue}{1.84} & \color{blue}{0.77}\\
    \hline
	\color{red}{\texttt{heuristic} size} & \color{red}{15446} & \color{red}{12392} & \color{red}{11548} & \color{red}{11676} & \color{red}{12794}\\
	\color{blue}{sr-\texttt{greedy} size} & \color{blue}{15395} & \color{blue}{12381} & \color{blue}{11661} & \color{blue}{12121} & \color{blue}{15189}\\
    sr-best size & 15394 & 12352 & 11512 & 11633 & 12741\\
        \bottomrule
    \end{tabularx}
\end{center}
}
    \caption{\label{tab:heur}
    Comparison of average solution sizes and average runtimes between the \texttt{heuristic} strategy, structural rounding with \texttt{greedy} lifting, and the best structural rounding result.
    }
\end{table}

\balance

%% file: sections/app-1.tex
\section{Complete Experiment Results}\label{app:one}

Here we give the complete results of our experiments, as described in Section~\ref{sec:experiments}. We split our data into subsections based on the parameter which was varied. Recall the color scheme we use for our data; red is the 4M expected edge data (procured), purple is the 4M expected edge data (prescribed), orange is the 40M expected edge data, and grey is the 200M expected edge data.

\subsection{Ratio of $|L|$ to $|R|$}\hfill\\

\begin{figure}[h!]
	\begin{centering}
	\vspace{-0.6cm}

	\includegraphics [width=0.85\textwidth]{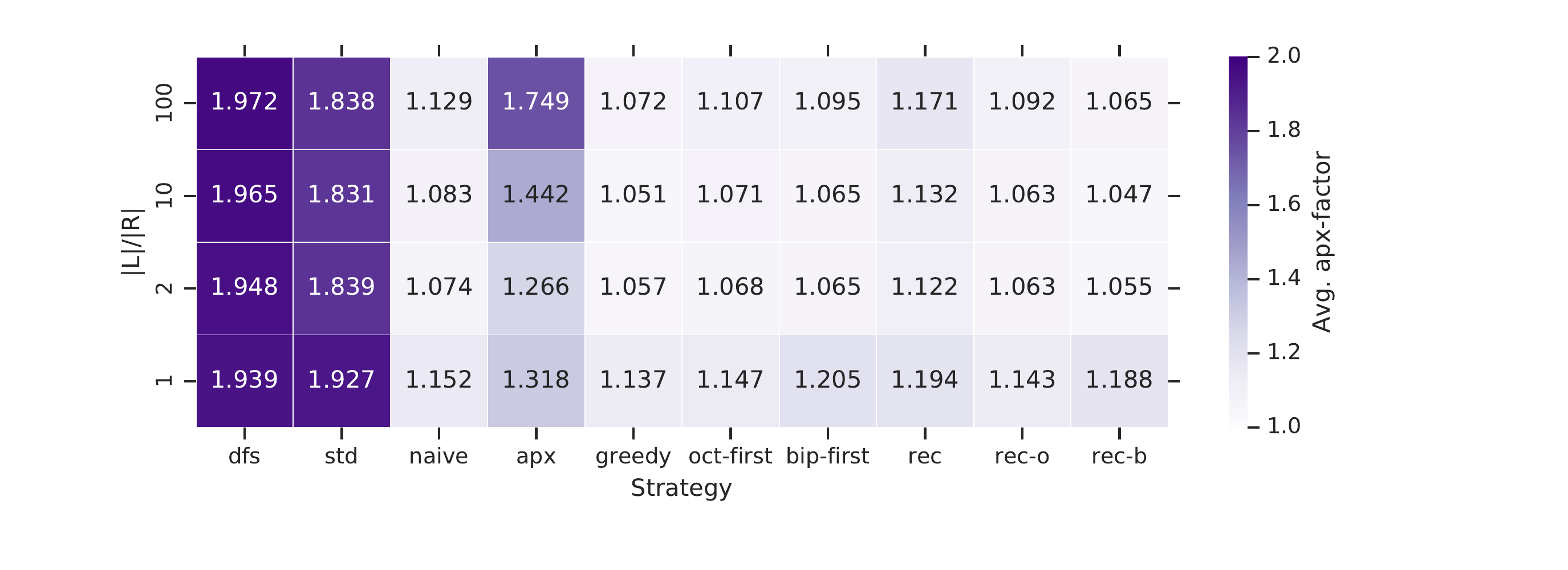}%

	\vspace{-0.6cm}

	\includegraphics [width=0.85\textwidth]{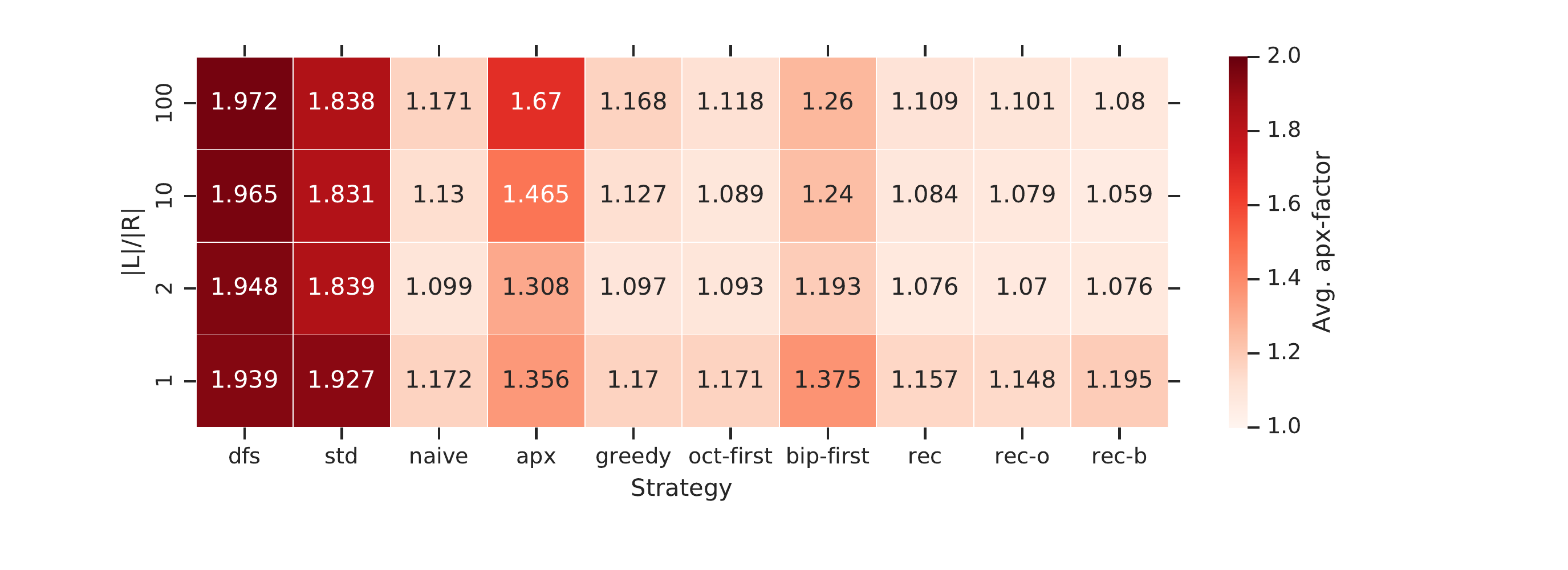}%

	\vspace{-0.6cm}

	\includegraphics [width=0.7\textwidth]{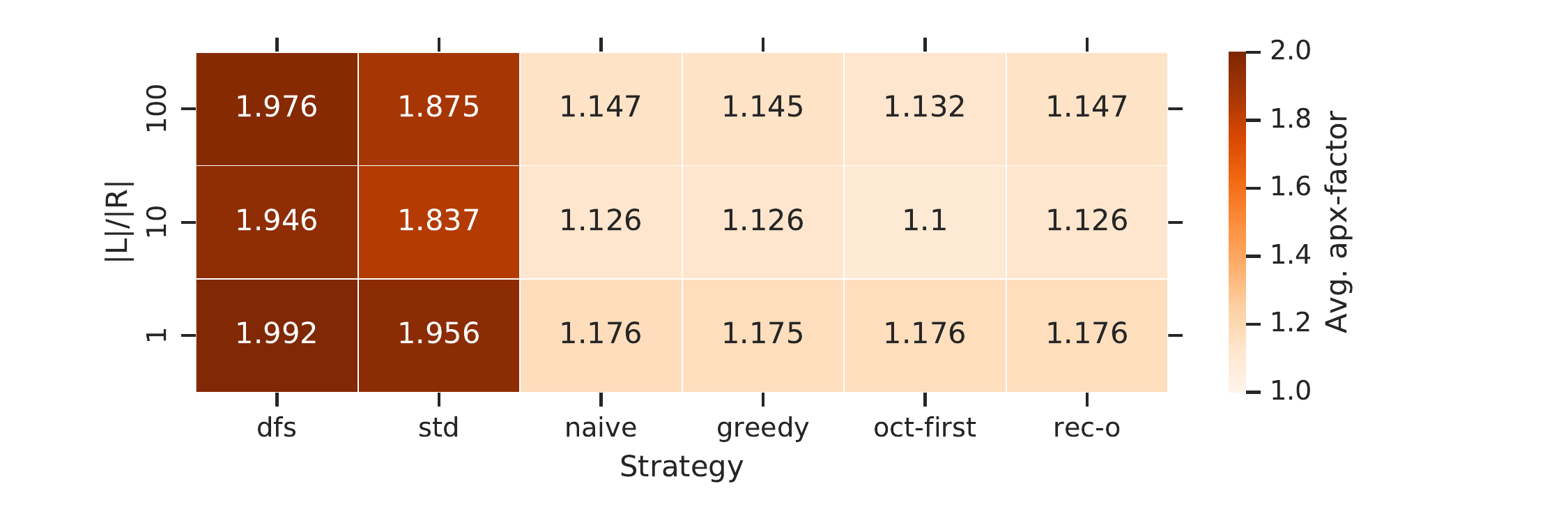}%

	\vspace{-0.3cm}

	\includegraphics [width=0.7\textwidth]{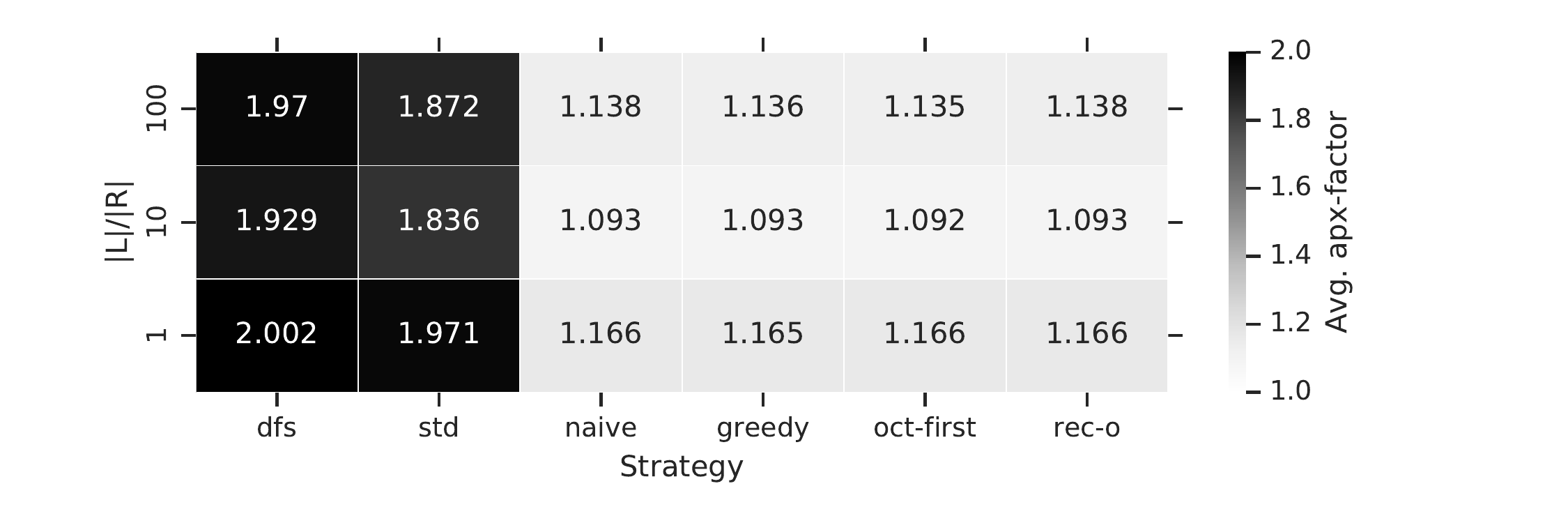}%

	\end{centering}
    \caption{\label{fig:bal}
The mean approximation ratios for each large synthetic expected edge setting, separated by $|L|/|R|$.
    }
\end{figure}

\clearpage

\subsection{Coefficient of variation between $O$ and $\{L,R\}$}\hfill\\

\begin{figure*}[h!]
	\begin{centering}
	\includegraphics [width=1\textwidth]{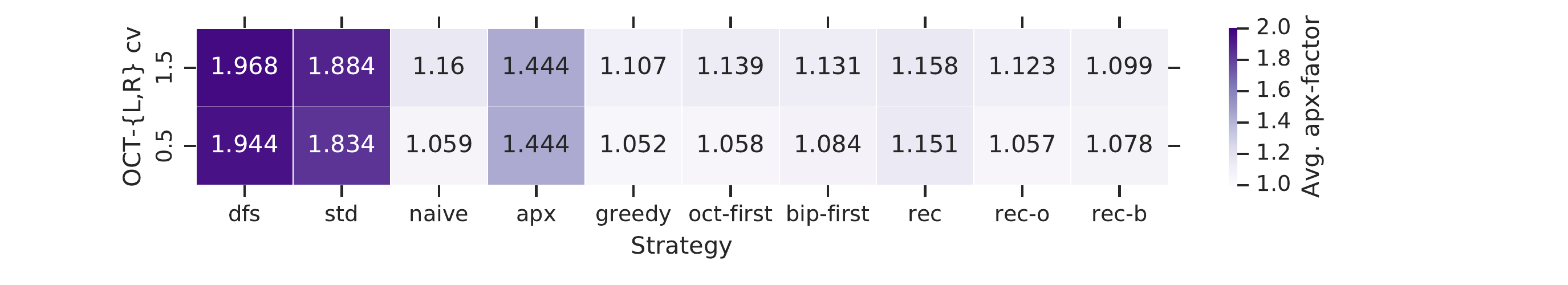}%

	\vspace{0.5cm}

	\includegraphics [width=1\textwidth]{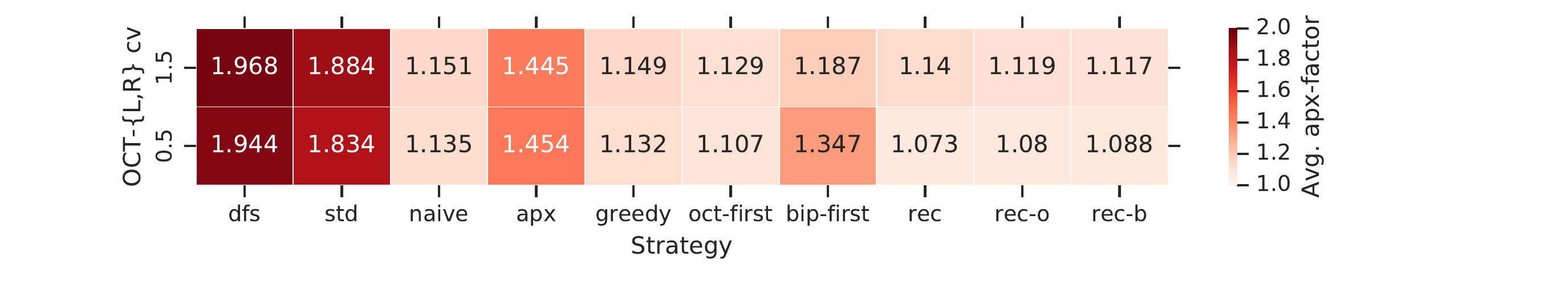}%

	\vspace{0.5cm}

	\includegraphics [width=0.8\textwidth]{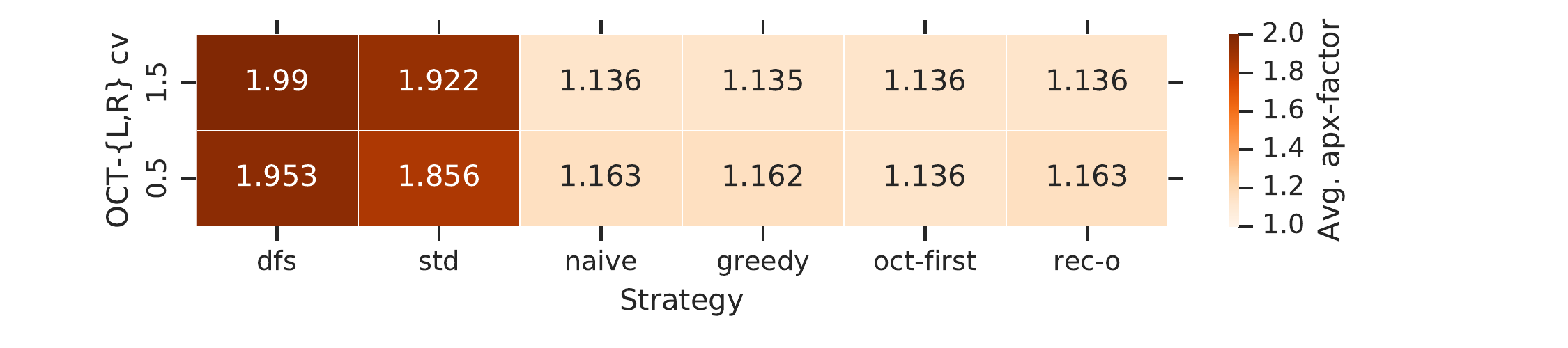}%

	\vspace{0.5cm}

	\includegraphics [width=0.8\textwidth]{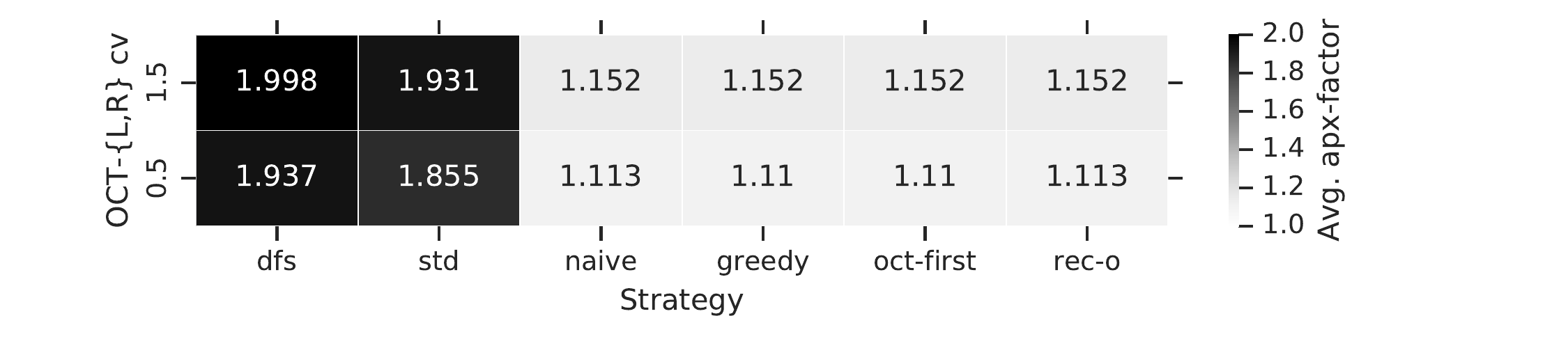}%

	\end{centering}

    \caption{\label{fig:cv}
The average approximation ratios over all synthetic graphs with 4-million expected edges in the prescribed setting, with 4-million expected edges in the procured, with 40-million expected edges, and with 200-million expected edges separated by the expected $cv$ between $O$ and $\{L,R\}$.
    }
\end{figure*}

\clearpage

\subsection{Ratio of OCT to $n$}\hfill\\

\begin{figure*}[h!]
	\begin{centering}
	\vspace{-0.65cm}

	\includegraphics [width=0.73\textwidth]{4M_pre_octsize-eps-converted-to.pdf}%

	\vspace{-0.85cm}

	\includegraphics [width=0.73\textwidth]{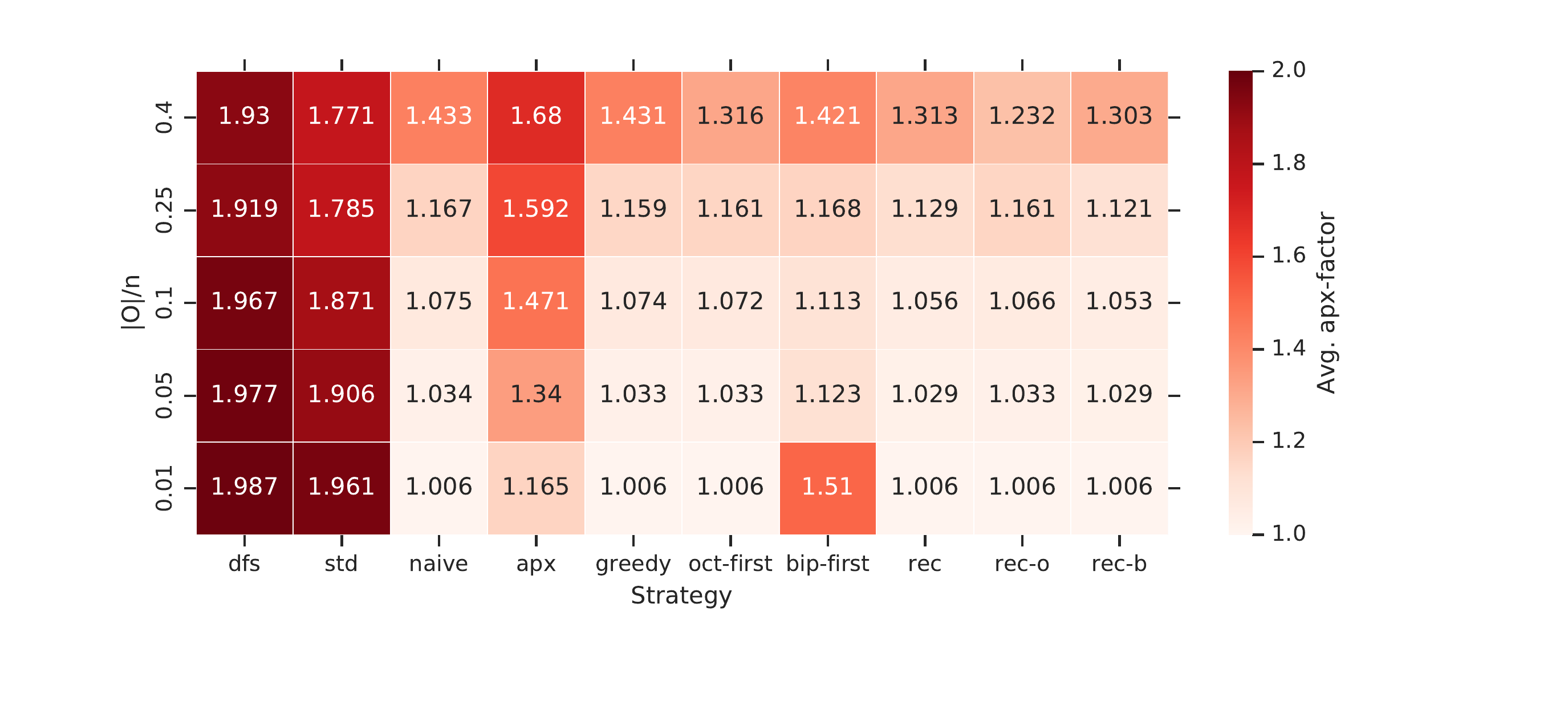}%

	\vspace{-0.85cm}

	\includegraphics [width=0.58\textwidth]{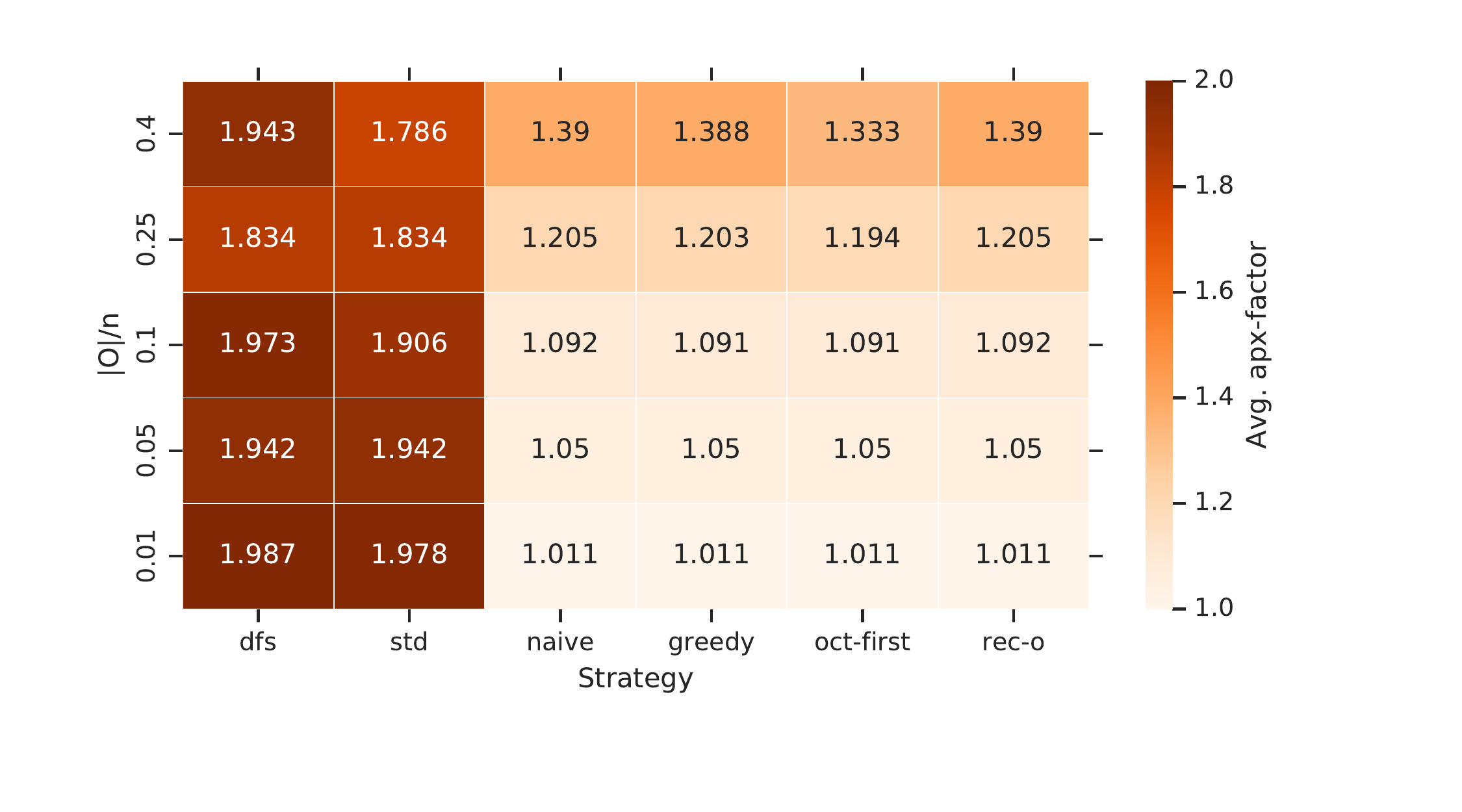}%

	\vspace{-0.8cm}

	\includegraphics [width=0.58\textwidth]{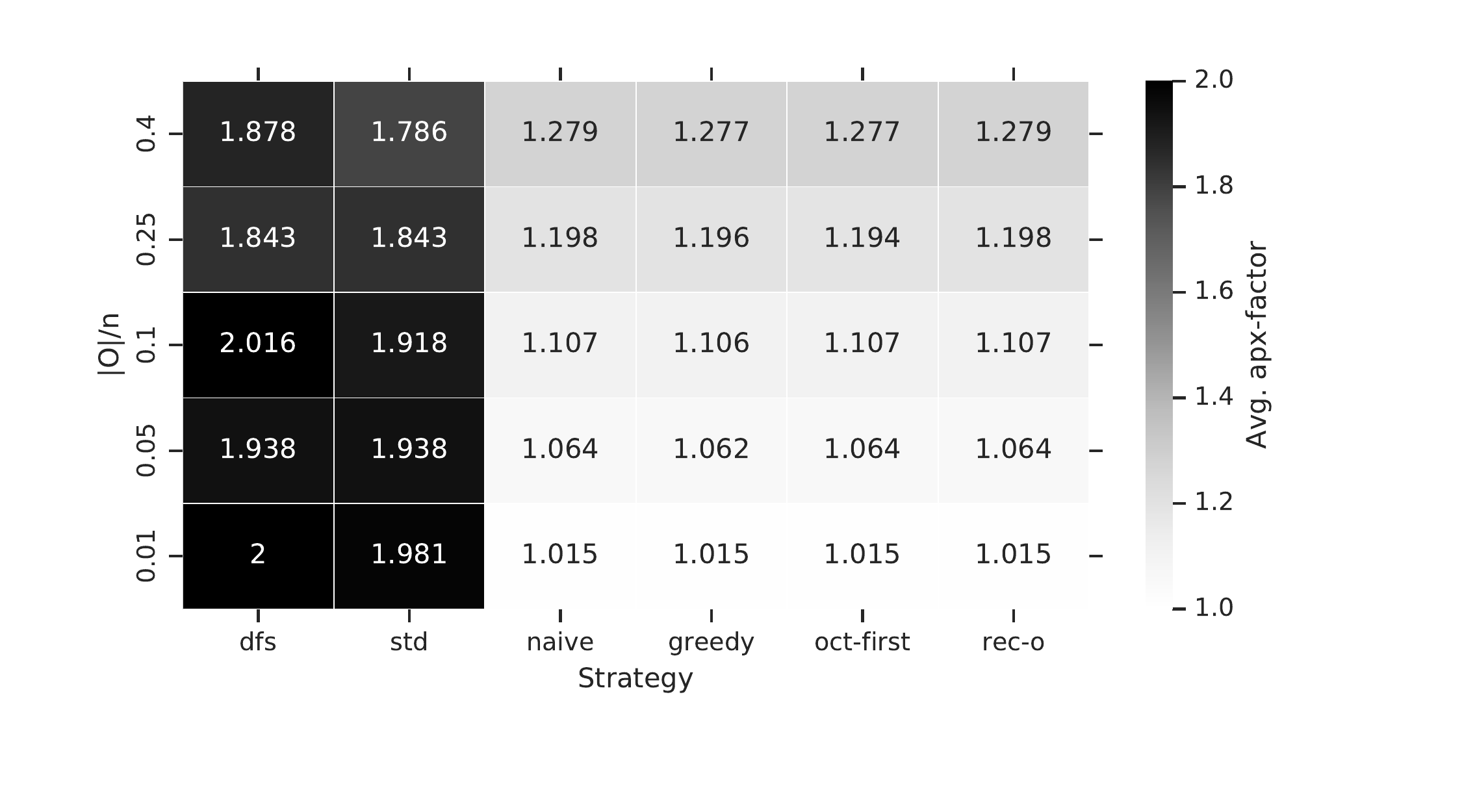}%

	\vspace{-0.8cm}

	\end{centering}

    \caption{\label{fig:octsize}
The average approximation ratios over all synthetic graphs with 4-million expected edges in the prescribed setting, with 4-million expected edges in the procured, with 40-million expected edges, and with 200-million expected edges separated by the proportion of the graph which is in $O$.
    }
\end{figure*}

\clearpage

\subsection{Expected edge density within OCT}\hfill\\

\begin{figure*}[h!]
	\begin{centering}
	\includegraphics [width=1\textwidth]{4M_pre_octoct-eps-converted-to.pdf}%

	\includegraphics [width=1\textwidth]{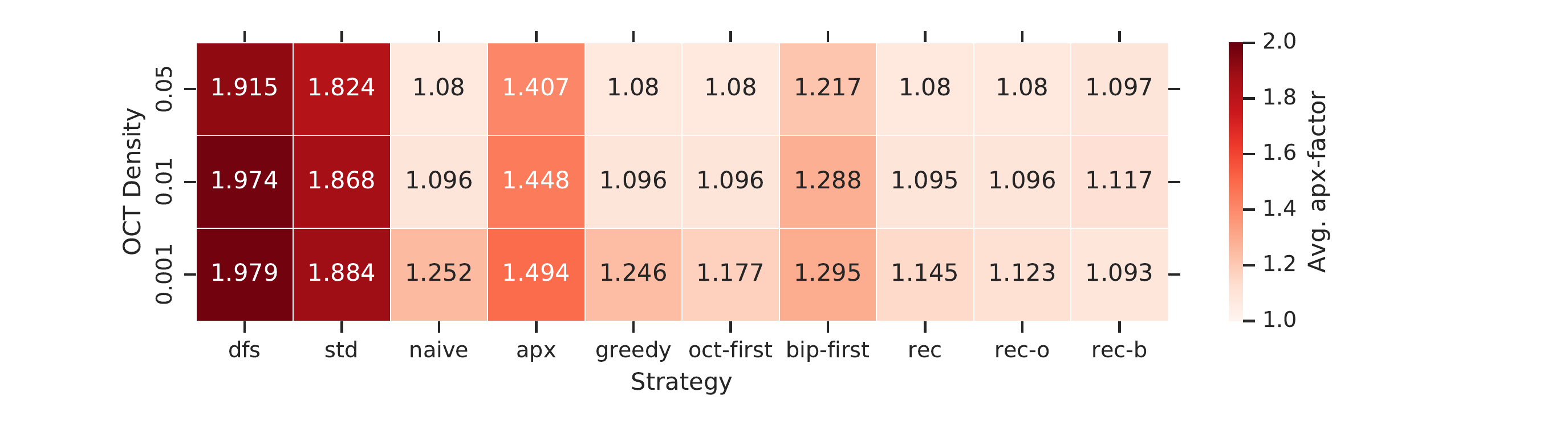}%

	\includegraphics [width=0.8\textwidth]{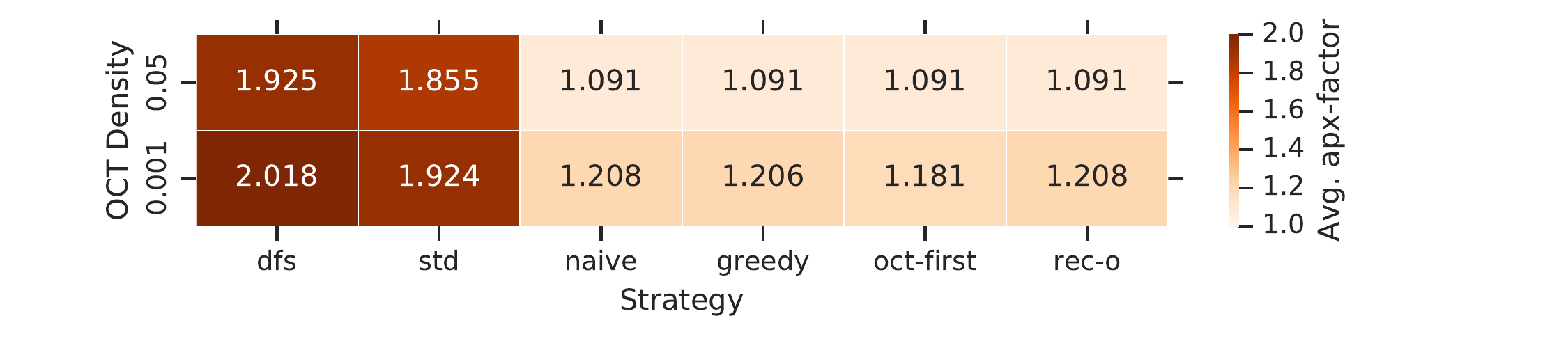}%

	\includegraphics [width=0.8\textwidth]{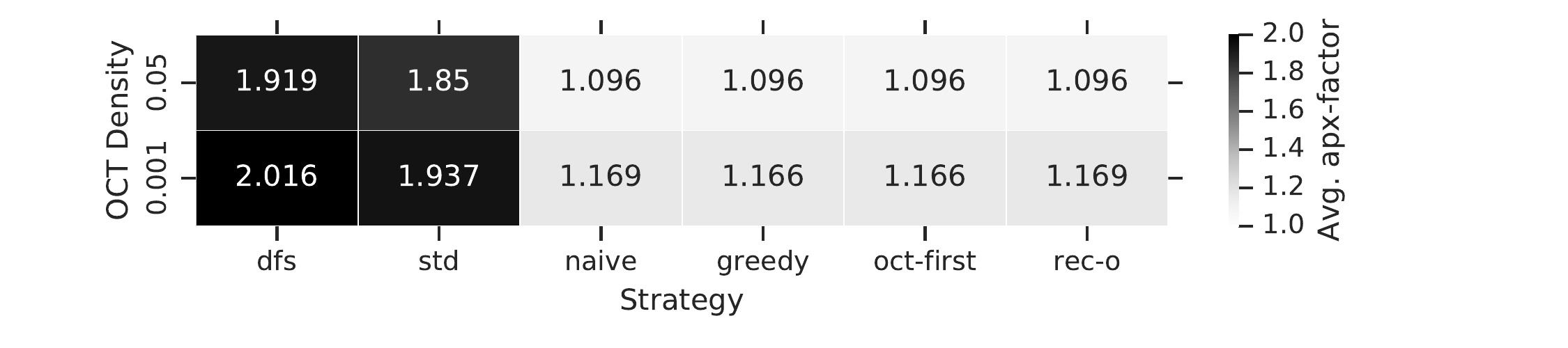}%

	\end{centering}

    \caption{\label{fig:octd}
The average approximation ratios over all synthetic graphs with 4-million expected edges in the prescribed setting, with 4-million expected edges in the procured, with 40-million expected edges, and with 200-million expected edges separated by the expected edge density within $O$.
    }
\end{figure*}

\clearpage

\subsection{Expected edge density between OCT and $\{L, R\}$}\hfill\\

\begin{figure*}[h!]
	\begin{centering}
	\includegraphics [width=1\textwidth]{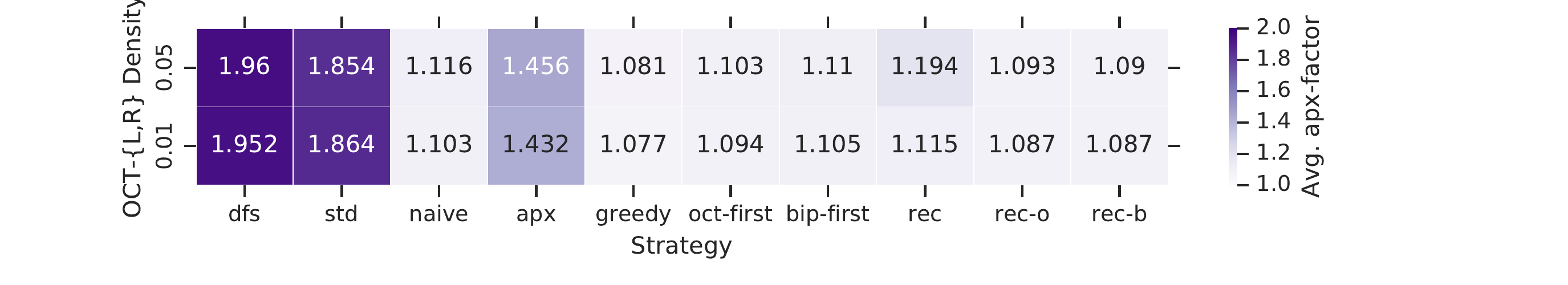}%

	\vspace{0.2cm}

	\includegraphics [width=1\textwidth]{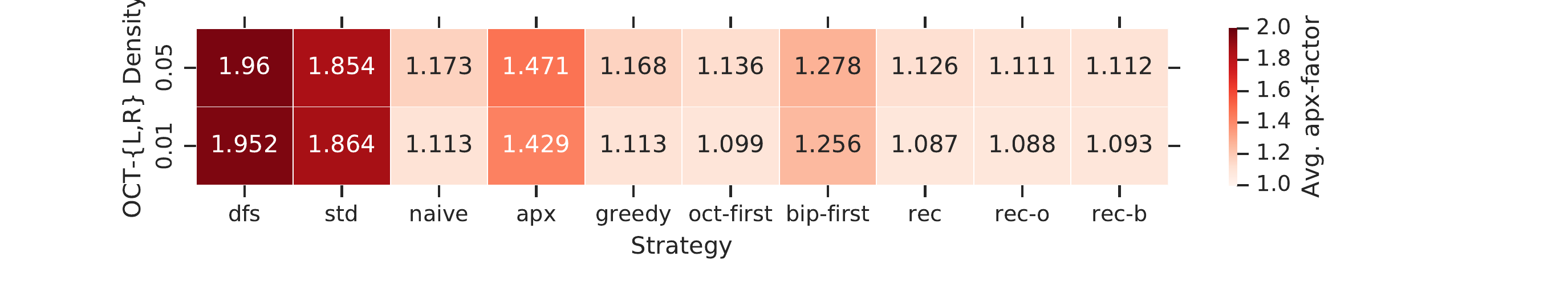}%

	\end{centering}
    \caption{\label{fig:old}
The average approximation ratios over all synthetic graphs with 4-million expected edges in the prescribed setting and in the procured separated by the expected edge density between $O$ and $\{L, R\}$.
    }
\end{figure*}

\subsection{Expected edge density between $L$ and $R$}\hfill\\

\begin{figure*}[h!]
	\begin{centering}
	\includegraphics [width=1\textwidth]{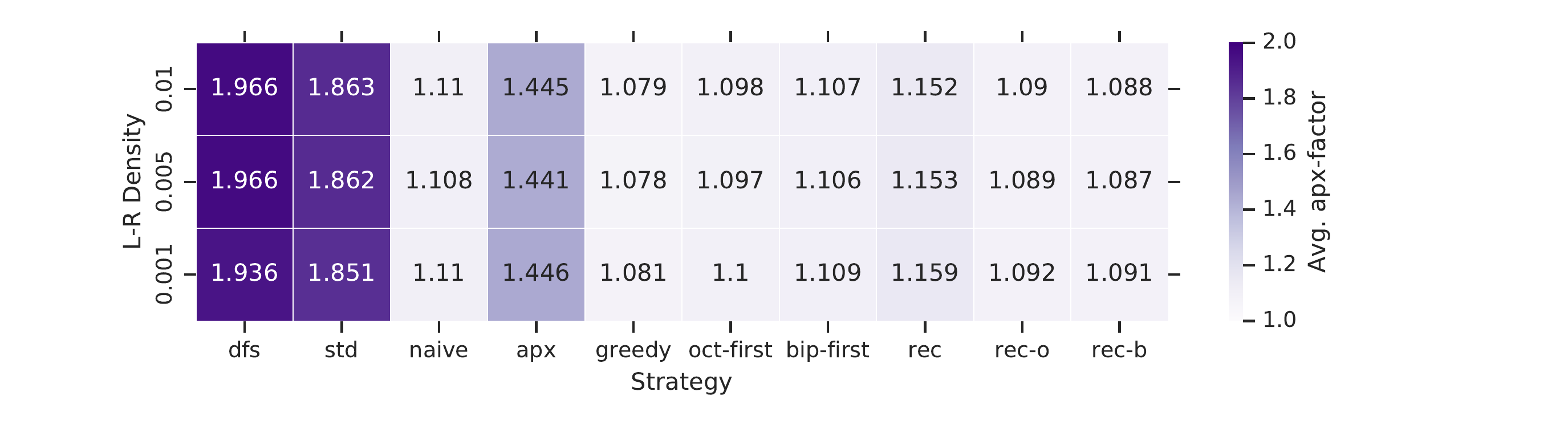}%

	\includegraphics [width=1\textwidth]{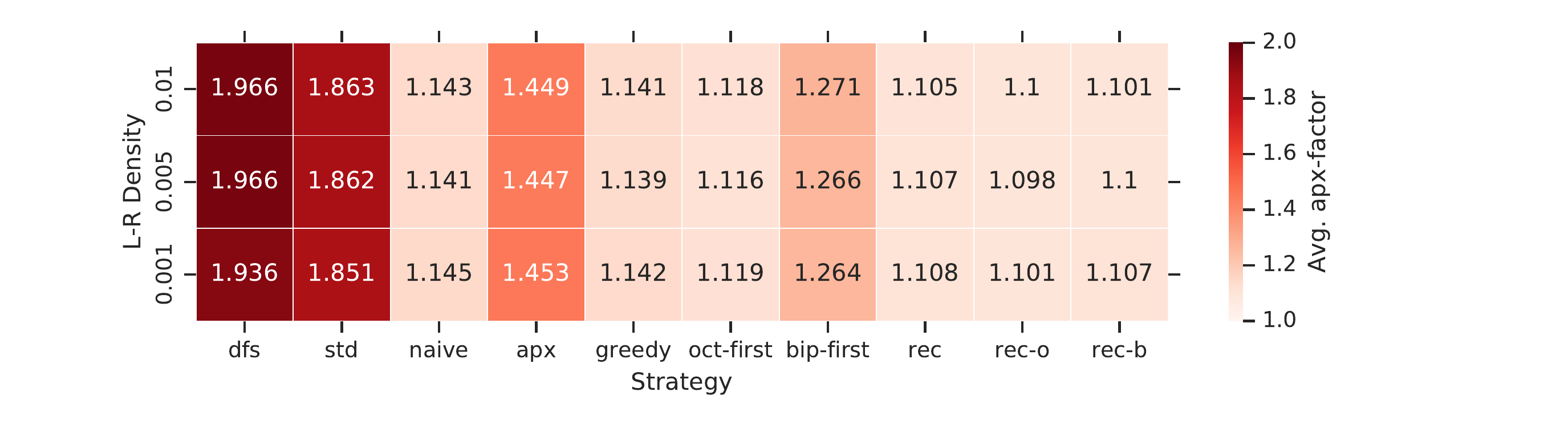}%

	\end{centering}
    \caption{\label{fig:lrd}
The average approximation ratios over all synthetic graphs with 4-million expected edges in the prescribed setting and in the procured setting separated by the expected edge density between $L$ and $R$.
    }
\end{figure*}

\clearpage

\section{Complete Variance Data}\label{app:var}\hfill\\

Here we include the complete variance data as mentioned in Section~\ref{sec:var}.\\

\begin{table*}[h!]
{\small
\begin{center}
    \begin{tabularx}{0.3\textwidth}{@{\extracolsep{\fill}} cc}
        \toprule
	module & avg. $\sigma^2/\mu$\\
	\midrule
	\texttt{DFS} 2-apx & 2.50e-05\\
	Std. 2-apx & 2.24e-05\\
	\midrule
	OCT-finding & 1.38e-04\\
	bipartite-solve & 7.86e-05\\
	\midrule
	\texttt{\naive} lift & 4.53e-07\\
	apx lift & 1.18e-05\\
	\texttt{greedy} lift & 7.55e-07\\
	\editfirst lift & 1.82e-04\\
	\bipfirst lift & 2.44e-04\\
	\rec lift & 3.86e-04\\
	\texttt{rec-oct} lift & 4.89e-04\\
	\texttt{rec-bip} lift & 2.83e-04\\
	\bottomrule
    \end{tabularx}
\end{center}
}
    \caption{\label{tab:variances}
We ran our framework 50 times on each graph in our corpus with 100000 expected edges, and for each component of the framework we computed the variance of its runtimes divided by its mean runtime for each graph.
We report the average of this value over all graphs for each component of the framework. 
    }
\end{table*}

\clearpage

\section{Procured setting: analyzing the effect of $|O|/n$ in generator}\label{app:pro}\hfill\\

Here we include figures for each of the five settings of $|O|/n$ in our generator, which capture the observed sizes of the OCT sets and the corresponding solution quality in the procured data. 

\begin{figure*}[h!]
	\begin{centering}

	\begin{minipage}{0.48\textwidth}
	\includegraphics [width=\textwidth]{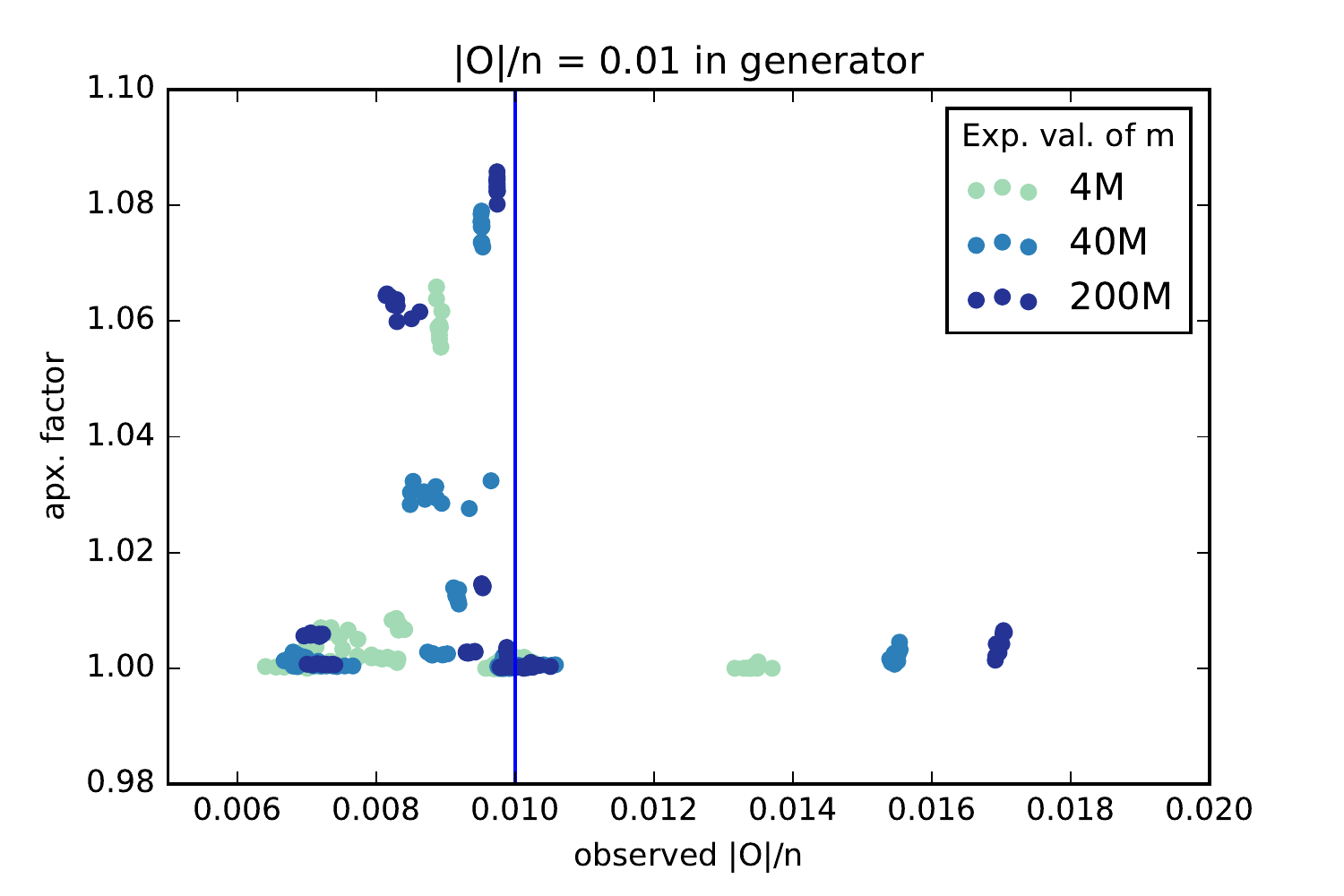}%

	\end{minipage}
	\begin{minipage}{0.48\textwidth}

	\includegraphics [width=\textwidth]{octsize005-eps-converted-to.pdf}%

	\end{minipage}

	\begin{minipage}{0.48\textwidth}
	\includegraphics [width=\textwidth]{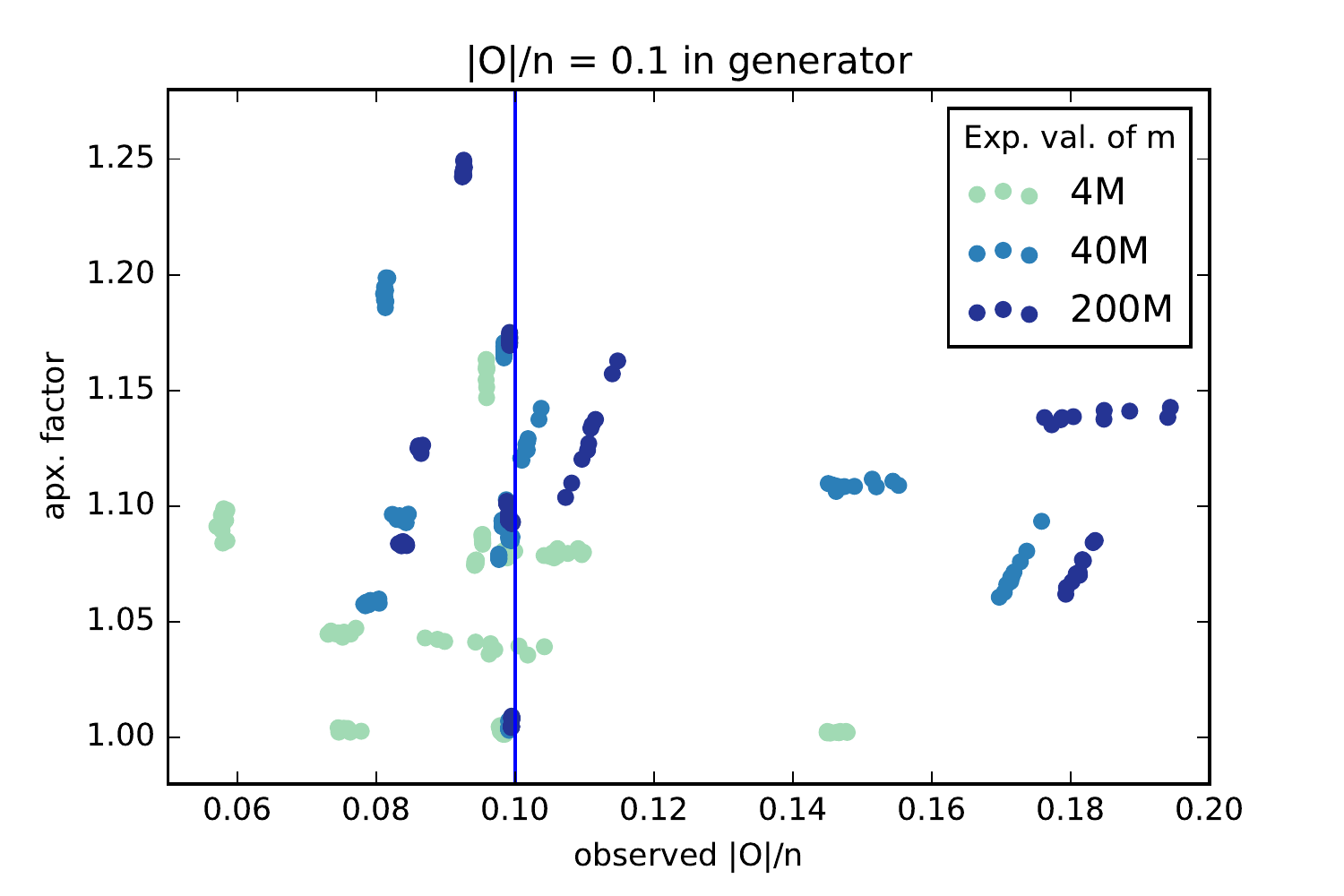}%

	\end{minipage}
	\begin{minipage}{0.48\textwidth}

	\includegraphics [width=\textwidth]{octsize025-eps-converted-to.pdf}%

	\end{minipage}

	\includegraphics [width=0.48\textwidth]{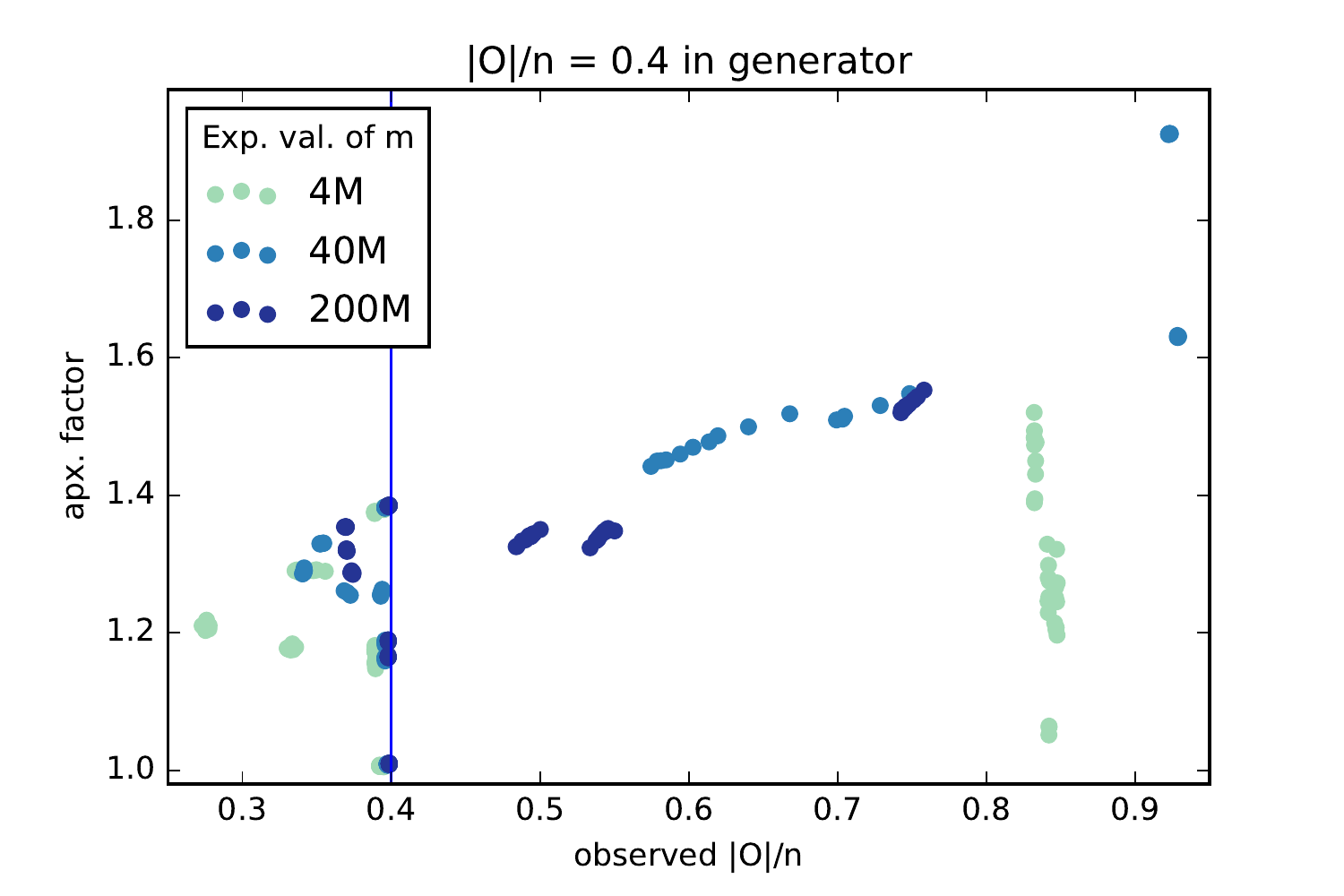}%

	\end{centering}
    \caption{\label{fig:octsizes}
		Correspondence between the size of the procured OCT set and solution quality across 4M-pro (green), 40M (blue), and 200M (red) corpuses; only graphs generated with parameters used in all experiments are included for consistency. The data is partitioned by prescribed relative OCT size $|O|/n$. We observe that performance is most consistent when the procured decomposition has OCT size close to the prescribed value. 
    }
\end{figure*}

%% file: main.bbl
\begin{thebibliography}{8}

\bibitem{AGARWAL}
P.~Agarwal, N.~Alon, B.~Aronov, and S.~Suri, {\em Can visibility graphs be represented compactly?}, Discrete \& Computational Geometry, 12 (1994), pp.~347--365.

\bibitem{AKIBA}
T.~Akiba and Y.~Iwata, {\em Branch-and-reduce exponential/fpt algorithms in practice: A case study of vertex cover}, Theoretical Computer Science, 609 (2016), pp.~211--225.

\bibitem{DEMAINE}
E.~D. Demaine, T.~D. Goodrich, K.~Kloster, B.~Lavallee, Q.~C. Liu, B.~D. Sullivan, A.~Vakilian, and A.~van der Poel, {\em Structural Rounding: Approximation Algorithms for Graphs Near an Algorithmically Tractable Class}, 27th Annual European Symposium on Algorithms (ESA 2019).



\bibitem{GOEMANS}
M.~X. Goemans and D.~P. ~Williamson, {\em Improved approximation algorithms for maximum cut and satisfiability problems using semidefinite programming}, Journal of the ACM (JACM), 42 (1995), pp.~1115--1145.

\bibitem{GOODRICH}
T.~Goodrich, E.~Horton, and B.~D. Sullivan, {\em Practical Graph Bipartization with Applications in Near-Term Quantum Computing}, arXiv preprint arXiv:1805.01041, 2018.

\bibitem{GUHA}
S.~Guha and S.~Khuller, {\em Approximation algorithms for connected dominating sets}, Algorithmica, 20 (1998), pp.~374--387.

\bibitem{GULPINAR}
N.~G{\"u}lpinar, G.~Gutin, G.~Mitra, and A.~Zverovitch, {\em Extracting pure network submatrices in linear programs using signed graphs}, Discrete Applied Mathematics, 137 (2004), pp.~359--372.


\bibitem{HOPCROFT}
J.~E. Hopcroft and R.~M. Karp, {\em An n\^{}5/2 algorithm for maximum matchings in bipartite graphs}, SIAM Journal on computing, 2 (1973), pp.~225--231.


\bibitem{HUFFNER}
F.~H{\"u}ffner, {\em Algorithm engineering for optimal graph bipartization}, International Workshop on Experimental and Efficient Algorithms, 2005, pp.~240--252.

\bibitem{IWATA}
Y.~Iwata, K.~Oka, and Y.~Yoshida, {\em Linear-time FPT algorithms via network flow}, SODA, 2014, pp.~1749--1761.

\bibitem{JOHNSON}
D.~S. ~Johnson, {\em Approximation algorithms for combinatorial problems}, Journal of computer and system sciences, 9 (1974), pp.~256--278.

\bibitem{KARAKOSTAS}
G.~Karakostas, {\em A better approximation ratio for the vertex cover problem}, International Colloquium on Automata, Languages, and Programming, (2005), pp.~1043--1050.

\bibitem{KARPINSKI}
M.~Karpinski and A.~Zelikovsky, {\em Approximating dense cases of covering problems}, (1996).

\bibitem{KONIG}
D.~K{\"o}nig, {\em Graphen und matrizen}, Mat. Fiz. Lapok, 38 (1931).

\bibitem{LAVALLEE}
B.~Lavallee, B.~D. Sullivan, and A.~van der Poel, Structural Rounding: Version v1.0, \url{http://doi.org/10.5281/zenodo.3401541}, September 2019.

\bibitem{LOKSHTANOV}
D.~Lokshtanov, S.~Saurab, and S.~Sikdar, {\em Simpler parameterized algorithm for OCT}, International Workshop on Combinatorial Algorithms, 2009, pp.~380--384.

\bibitem{PANCONESI}
A.~Panconesi and M.~Sozio, {\em Fast hare: A fast heuristic for single individual SNP haplotype reconstruction}, International workshop on algorithms in bioinformatics, 2004, pp.~266--277.

\bibitem{PAPADIMITROU}
C.~H. Papadimitrou and K.~Steiglitz, {\em Combinatorial optimization: algorithms and complexity}, (1982).

\bibitem{GRAPHS}
\textsf{network-corpus}, \url{github.com/microgravitas/network-corpus}, February 2018.

\bibitem{SAVAGE}
C.~Savage, {\em Depth-first search and the vertex cover problem}, Information Processing Letters, 14 (1982), pp.~233--235.

\bibitem{SCHROOK}
J.~Schrook, A.~McCaskey, K.~Hamilton, T.~Humble, and N.~Imam, {\em Recall Performance for Content-Addressable Memory Using Adiabatic Quantum Optimization} Entropy, 19 (2017).

\bibitem{VAZIRANI}
V.~V. Vazirani, {\em Approximation algorithms}, 2013.

\bibitem{ZHANG}
Y.~Zhang, C.~A. Phillips, G.~L. Rogers, E.~J. Baker, E.~J. Chesler, and M.~A. Langston, {\em On finding bicliques in bipartite graphs: a novel algorithm and its application to the integration of diverse biological data types}, BMC Bioinformatics, 15 (2014).

\end{thebibliography}
